\documentstyle[aps,preprint,prl,epsfig]{revtex}
\begin{document}
\title{The interacting boson model with SU(3) charge symmetry\\
and its application to even--even N$\approx$Z nuclei}
\author{J.E.~Garc\'{\i}a--Ramos$^1$ and P.~Van Isacker$^2$}
\address{
$^1$ Departamento de F\'{\i}sica At\'{o}mica, Molecular y Nuclear,
Universidad de Sevilla, Apartado 1065, 41080 Sevilla, Spain}
\address{
$^2$ Grand Acc\'el\'erateur National d'Ions Lourds, B.P.~5027, 
F-14076 Caen Cedex 5, France}
\date{\today}
\maketitle

\begin{abstract}
The isospin-invariant interacting boson model IBM-3
is analyzed in situations
where $SU_T(3)$ charge symmetry
[or, equivalently, $U_L(6)$ $sd$ symmetry]
is conserved.
Analytic expressions for energies,
electromagnetic transitions, 
two-nucleon transfer probabilities,
and boson-number expectation values are obtained
for the three possible dynamical symmetry limits,
$U(5)$, $SU(3)$, and $O(6)$.
Results found in IBM-3
are related to corresponding ones in IBM-1 and IBM-2.
Numerical calculations are presented for $f_{7/2}$-shell nuclei
and some features that distinguish IBM-3
from its predecessors IBM-1 and IBM-2 are pointed out.
\end{abstract}

\pacs{PACS numbers: 21.60.-n, 21.60.Fw}

\newpage
\section{Introduction}
The interacting boson model (IBM) \cite{Arim75,Iach87}
was originally proposed as a phenomenological model
couched in terms of $s$ and $d$ bosons
of which the microscopic structure (e.g., in terms of the shell model)
was unknown.
This state of affairs quickly changed
with the realization that the bosons 
could be interpreted as correlated (Cooper) pairs
with angular momentum $J=0$ and $J=2$,
formed by the nucleons in the valence shell.
A connection with an underlying shell-model picture
could be established by making a distinction
between a proton pair ($\pi$ boson) and a neutron pair ($\nu$ boson),
the resulting model being referred to
as the proton--neutron interacting boson model
or IBM-2\cite{Arim77,Otsu78,Otsu81}.

The IBM-2 has been applied extensively and successfully
to even--even medium-mass and heavy nuclei \cite{Iach87}
where the protons and neutrons occupy different valence shells.
In the latter situation it is natural
to assume correlated proton--proton and neutron--neutron pairs,
and to include (longe-range) proton--neutron correlations
through a quadrupole interaction in the Hamiltonian.
In lighter $N\approx Z$ nuclei
where the protons and the neutrons occupy the same valence shell,
this approach no longer is valid
since there is no reason not to include proton--neutron $T=1$ pairs
in such nuclei.
The inclusion of proton--neutron $T=1$ pairs ($\delta$ bosons)
has indeed been proposed by Elliott and White \cite{Elli80}
and the resulting model has been named IBM-3.
Since the IBM-3 contains a complete isospin triplet of $T=1$ bosons,
it is possible to construct IBM-3 Hamiltonians
that conserve isospin symmetry
and this feature can be exploited
to establish a more direct correspondence
between the IBM-3 and the shell model \cite{Evan85}.
As a final refinement of the interacting boson model,
Elliott and Evans \cite{Elli81}
proposed the inclusion of a proton--neutron $T=0$ pair ($\sigma$ boson)
leading to a version referred to as IBM-4.
Both $T=0$ and $T=1$ proton--neutron ($T_z=0$) bosons
play an important role in $N=Z$ nuclei
but their importance decreases
with increasing difference $|N-Z|$.
Furthermore, $T=0$ bosons are crucial in odd--odd nuclei
while even--even nuclei seem to be adequately described
with $T=1$ bosons only.
These observations then define the scope of the present work:
the IBM-3 is geared towards applications to even--even $N\approx Z$ nuclei
although its results can be extended to isobaric analog states
in neighboring odd--odd nuclei.

The IBM-3 has a rich algebraic structure
that starts from the dynamical algebra $U(18)$
and allows several dynamical symmetries containing the $O_T(3)$ subalgebra,
necessary to conserve isospin symmetry.
In this paper the dynamical symmetries are analyzed
that arise after the reduction of $U(18)$
to the direct product $U_L(6)\otimes SU_T(3)$
and thus assume a separation of the orbital ($sd$) and isospin sectors.
The latter approximation plays the same role in IBM-3
as does the assumption of $F$-spin symmetry in IBM-2 \cite{Arim77}.
In realistic IBM-2 calculations
the lowest-energy states are (approximately) symmetric
under the exchange of proton and neutron indices,
that is, they have a large component
in the symmetric representation $[N]$ of $U_L(6)$
(or, equivalently, a large component with maximal $F$ spin,
$F_{\rm max}=N/2$).
Non-symmetric states $[N-1,1]$ occur at higher energies \cite{Bohl84}.
Although realistic IBM-2 Hamiltonians
may contain important $F$-spin mixing interactions,
$F$ spin is usually an approximately conserved symmetry
because of the existence of a relatively large Majorana interaction.
The Majorana operator is diagonal in $U_L(6)$ [or $SU_F(2)$]
and separates the different $U_L(6)$ representations $[N],[N-1,1],\dots$
[or $SU_F(2)$ representations with $F=F_{\rm max},F_{\rm max}-1,\dots$].
As a result, the orbital and charge (or $F$-spin) spaces
are approximately decoupled in IBM-2.
The situation is very similar in IBM-3,
the $U_L(6)\otimes SU_F(2)$ of IBM-2 being replaced by
$U_L(6)\otimes SU_T(3)$,
which by analogy can be expected to be an approximate symmetry algebra.
This is indeed confirmed in some realistic IBM-3 calculations
for $pf$-shell nuclei \cite{Evan85,Abde88,Abde89,Elli96}
which show that, at low energy,
states can be approximately classified by $U_L(6)$ representations
$[N]$ or $[N-1,1]$.
Representations of lower symmetry such as $[N-2,2]$
are less well realized
but these are not considered in this paper.

The structure of this paper is as follows.
First, in Section~\ref{sec-al-struc},
a general overview of the algebraic structure of IBM-3 is given;
in Section~\ref{sec-ham} the IBM-3 Hamiltonian is specified
in different representations.
In Section~\ref{sec-dyn-sym} a brief review is given
of the three symmetries that are analyzed in this paper.
Sections~\ref{sec-elec-tran} and \ref{sec-two-nucl}
are devoted to the analysis of
electromagnetic  transitions and
two-nucleon transfer probabilities,
respectively,
and in Section~\ref{sec-bos-num}
analytic expressions are given for boson-number expectation values.
In Section~\ref{sec-exp}
the most relevant predictions of IBM-3 (as contrasted with IBM-1 and IBM-2)
are pointed out
and results of numerical calculations in the $f_{7/2}$ shell are shown.
Finally, conclusions are presented in Section~\ref{sec-conclu}.

\section{Algebraic structure}
\label{sec-al-struc}
The basic building blocks of the IBM-3
are assigned the orbital angular momenta $l=0$ and $l=2$
($s$ and $d$ bosons)
and isospin $T=1$ with isospin projection $\mu=+1,0,-1$
for the $\pi$, $\delta$, and $\nu$ boson
(the proton--proton, neutron--proton, and neutron--neutron pairs), 
respectively.
The corresponding creation and annihilation operators can be written as
\begin{equation}
b^\dag_{lm,1\mu},
\qquad
b_{lm,1\mu}.
\label{Bos}
\end{equation}
These operators are assumed
to satisfy the customary boson commutation relations
\begin{equation}
[b_{lm,1\mu},b^\dag_{l'm',1\mu'}]=
\delta_{ll'}\delta_{mm'}\delta_{\mu\mu'},
\qquad
[b^\dag_{lm,1\mu},b^\dag_{l'm',1\mu'}]=
[b_{lm,1\mu},b_{l'm',1\mu'}]=0.
\label{ComRel}
\end{equation}
With the operators (\ref{Bos}) 324 bilinear combinations,
\begin{equation}
b^\dag_{lm,1\mu}b_{l'm',1\mu'},
\label{GenUnc}
\end{equation}
can be constructed that generate the algebra $U(18)$,
as can be shown explicitly from the commutation relations
\begin{eqnarray}
\lefteqn{[b^\dag_{lm,1\mu}b_{l'm',1\mu'},
b^\dag_{l''m'',1\mu''}b_{l'''m''',1\mu'''}]}
\nonumber\\
&=&
b^\dag_{lm,1\mu}b_{l'''m''',1\mu'''}
\delta_{l'l''}\delta_{m'm''}\delta_{\mu'\mu''}-
b^\dag_{l''m'',1\mu''}b_{l'm',1\mu'}
\delta_{ll'''}\delta_{mm'''}\delta_{\mu\mu'''}.
\label{ComRelUnc}
\end{eqnarray}
The operators of all physical observables
will be expressed in terms of the generators (\ref{GenUnc})
and, consequently, the dynamical algebra of IBM-3 is $U(18)$.

Two important invariances occur in the context of IBM-3:
rotational and isospin invariance.
The first one is an exact symmetry
and leads to conservation of total angular momentum, here denoted as $L$;
the second gives rise to the isospin quantum number $T$.
Isospin is only an approximate symmetry
(mainly broken by the Coulomb interaction)
but throughout this paper it is assumed to be exact.
Given these two invariances it is convenient
to take combinations of the operators (\ref{GenUnc})
that have definite transformation properties
under rotations in physical and isospin space.
The generators of $U(18)$ can thus also be written as
\begin{equation}
(b^\dag_{l,1}\times\tilde{b}_{l',1})^{(L,T)}_{M_L,M_T}=
\sum_{mm'\mu\mu'}
\langle lm\;l'm'|LM_L\rangle
\langle 1\mu\;1\mu'|TM_T\rangle
b^\dag_{lm,1\mu}\tilde{b}_{l'm',1\mu'},
\label{GenCou}
\end{equation}
where the symbol between angle brackets is a Clebsch--Gordan coefficient
and $\tilde{b}_{lm,1\mu}\equiv(-)^{l-m+1-\mu}b_{l-m,1-\mu}$
has the appropriate transformation properties
under rotations in physical and isospin space.
For completeness we also give the commutation relations
among the coupled generators of $U(18)$:
\begin{eqnarray}
\lefteqn{[(b^\dag_{l,1}\times\tilde{b}_{l',1})^{(L,T)}_{M_L,M_T},
          (b^\dag_{l'',1}\times\tilde{b}_{l''',1})^{(L',T')}_{M'_L,M'_T}]}
\nonumber\\
&=&
\sum_{L''M''_LT''M''_T}
\hat{L}\hat{L}'\hat{T}\hat{T}'
\langle LM_L\;L'M'_L|L''M''_L\rangle
\langle TM_T\;T'M'_T|T''M''_T\rangle
\nonumber\\
&\times&\left[
(-)^{L''+T''}
\left\{\begin{array}{ccc}
L&L'&L''\\
l'''&l&l'
\end{array}\right\}
\left\{\begin{array}{ccc}
T&T'&T''\\
1&1&1
\end{array}\right\}
\delta_{l'l''}
(b^\dag_{l,1}\times\tilde{b}_{l''',1})^{(L'',T'')}_{M''_L,M''_T}
\right.\nonumber\\
&&\left.
-(-)^{L+L'+T+T'}
\left\{\begin{array}{ccc}
L&L'&L''\\
l''&l'&l
\end{array}\right\}
\left\{\begin{array}{ccc}
T&T'&T''\\
1&1&1
\end{array}\right\}
\delta_{ll'''}
(b^\dag_{l'',1}\times\tilde{b}_{l',1})^{(L'',T'')}_{M''_L,M''_T}
\right],
\label{ComRelCou}
\end{eqnarray}
where $\hat{L}\equiv\sqrt{2L+1}$
and the symbol between curly brackets is a Racah coefficient.

The dynamical algebra $U(18)$ has a rich substructure.
One is, however, not interested
in all possible algebraic decompositions of $U(18)$
but only in those that conserve angular momentum and isospin,
that is, the ones containing the angular momentum algebra $O_L(3)$
and the isospin algebra $O_T(3)$.
A possible way to impose these symmetries
is to consider the reduction
\begin{equation}
\begin{array}{ccccc}
U(18)&\supset&U_L(6)&\otimes&SU_T(3)\\
\downarrow&&\downarrow&&\downarrow\\
{[N]}&&[N_1,N_2,N_3]&&(\lambda_T,\mu_T)
\end{array},
\label{Red}
\end{equation}
which corresponds to a decomposition of states
into an orbital (or $sd$) and an isospin part.
Because of the overall symmetry in $U(18)$,
the $U_L(6)$ and $SU_T(3)$ representations are the same
(i.e., they correspond to the same Young diagram)
and this leads to $U_L(6)$ representations $[N_1,N_2,N_3]$
that can have up to three rows with length $N_i$ and with $N_1+N_2+N_3=N$.
This situation should be compared with IBM-2
where at most two-rowed representations can occur in $U(6)$
and IBM-1
which only contains symmetric (one-rowed) $U(6)$ representations.
In (\ref{Red}) Elliott's $SU(3)$ labels are used \cite{Elli58}
which are related to the usual row labels
by $\lambda_T=N_1-N_2$ and $\mu_T=N_2-N_3$.

The generators of $U_L(6)$ and $SU_T(3)$ are obtained
by contracting in the isospin and orbital indices, respectively.
The following coupled form of the generators results:
\begin{eqnarray}
U_L(6)&:&(b^\dag_{l,1}\times\tilde{b}_{l',1})^{(L,0)}_{M_L,0},
\nonumber\\
U_T(3)&:&
\sum_l\sqrt{2l+1}(b^\dag_{l,1}\times\tilde{b}_{l,1})^{(0,T)}_{0,M_T},
\label{GenU6U3}
\end{eqnarray}
with $l,l'=0,2$,
and $L$ and $T$ running over all values
compatible with angular momentum coupling.
The $SU_T(3)$ algebra consists of the generators
of $U_T(3)$ with $T=1$ and $T=2$.

The classification (\ref{Red}) is {\em sufficient} to ensure
invariance under rotations in physical and isospin space.
It is, however, not a {\em necessary} condition
and classes of Hamiltonians have been shown to exist \cite{Gino96}
that conserve angular momentum and isospin
but do not proceed via the reduction (\ref{Red}).
In this paper the latter reductions are not considered
but only those that conserve $SU_T(3)$
which can be considered as a charge symmetry algebra.
Note that $SU_T(3)$
is not a fundamental symmetry such as angular momentum or isospin,
and that it may well be broken by specific boson--boson interactions.
The requirement of $SU_T(3)$ charge symmetry
obviously restricts the applicability of the results derived here.
On the other hand,
it must be emphasized that $SU_T(3)$ is equivalent to $U_L(6)$
which is a symmetry of basic importance in the IBM.

\section{The IBM-3 Hamiltonian}
\label{sec-ham}
Any Hamiltonian
which is invariant under rotations in physical and isospin space
can be written as $L$- and $T$-scalar combinations
of the generators (\ref{GenCou}).
If up to two-body interactions in the bosons are taken,
the most general Hamiltonian is
\begin{eqnarray}
\label{HamSta}
\hat{H}&=&
\sum_l
\epsilon_l
\sqrt{3(2l+1)}
(b^\dag_{l,1}\times\tilde{b}_{l,1})^{(0,0)}_{0,0}
\\
&+&
\sum_{{l_1l_2l'_1l'_2}\atop{LT}}
v_{l_1l_2l'_1l'_2}^{LT}
\sqrt{{(2L+1)(2T+1)}\over{(1+\delta_{l_1l_2})(1+\delta_{l'_1l'_2})}}
\Big((b^\dag_{l_1,1}\times b^\dag_{l_2,1})^{(L,T)}\times
(\tilde{b}_{l'_1,1}\times\tilde{b}_{l'_2,1})^{(L,T)}\Big)^{(0,0)}_{0,0}.
\nonumber
\end{eqnarray}
The coefficients $\epsilon_0$ and $\epsilon_2$
are the $s$- and $d$-boson energies,
which by virtue of isospin invariance
are independent of the nature of the bosons ($\pi$, $\delta$, or $\nu$).
The coefficients  $v_{l_1l_2l'_1l'_2}^{LT}$ are the interaction 
matrix elements between normalized two-boson states, 
\begin{equation}
v_{l_1l_2l'_1l'_2}^{LT}
\equiv
\langle l_1l_2;LT|\hat{H}|l'_1l'_2;LT\rangle.
\label{TwoBosMat}
\end{equation}
The form (\ref{HamSta}) is referred to as
the standard representation of the IBM-3 Hamiltonian.

The IBM-3 Hamiltonian alternatively
can be written in a multipole expansion form as
\begin{equation}
\hat{H}_{\rm mul}=
\sum_l
\eta_l
\sqrt{3}\hat{l}
(b^\dag_{l,1}\times\tilde{b}_{l,1})^{(0,0)}_{0,0}
+
\sum_{LT}
\kappa_{LT}
\hat{L}\hat{T}
(\hat{T}^{(L,T)}\times\hat{T}^{(L,T)})^{(0,0)}_{0,0},
\label{HamMul}
\end{equation}
where $\hat{T}^{(L,T)}$
are isoscalar ($T=0$), isovector ($T=1$), and isotensor ($T=2$)
multipole operators of the form
\begin{equation}
\hat{T}^{(L,T)}_{M_L,M_T}=
\sum_{l_1l_2}
\chi_{l_1l_2}^{LT}
(b^\dag_{l_1,1}\times\tilde{b}_{l_2,1})^{(L,T)}_{M_L,M_T}.
\label{OpeMul}
\end{equation}
The last term in (\ref{HamMul})
is not a pure two-body term but it contains one-body pieces as well
and therefore the parameters $\eta_l$ in (\ref{HamMul})
do not coincide with the single-boson energies $\epsilon_l$
of (\ref{HamSta}).

The standard and multipole forms (\ref{HamSta}) and (\ref{HamMul})
are two equivalent ways of writing the IBM-3 Hamiltonian,
which can be related as follows:
\begin{eqnarray}
\epsilon_l&=&
\eta_l+
\sum_{l'LT}
(-)^{L+T}
{{(2L+1)(2T+1)}\over{3(2l+1)}}
\kappa_{LT}\chi_{ll'}^{LT}\chi_{l'l}^{LT},
\nonumber\\
v_{l_1l_2l'_1l'_2}^{LT}&=&
\sum_{L'T'}
(-)^{L+T+L'+T'}
{{2(2L'+1)(2T'+1)}\over{(1+\delta_{l_1l_2})(1+\delta_{l'_1l'_2})}}
\kappa_{L'T'}
\left\{\begin{array}{ccc}
1&1&T\\
1&1&T'
\end{array}\right\}
\nonumber\\
&\times&\left[
\chi_{l_1l'_1}^{L'T'}
\chi_{l_2l'_2}^{L'T'}
\left\{\begin{array}{ccc}
l_1&l_2&L\\
l'_2&l'_1&L'
\end{array}\right\}
+(-)^{L+T}
\chi_{l_1l'_2}^{L'T'}
\chi_{l_2l'_1}^{L'T'}
\left\{\begin{array}{ccc}
l_1&l_2&L\\
l'_1&l'_2&L'
\end{array}\right\}
\right].
\label{MulSta}
\end{eqnarray}

A third way of parametrizing the IBM-3 Hamiltonian exists
which relies on the algebraic (sub)structure of $U_L(6)\otimes SU_T(3)$
and the associated Casimir operators.
The starting point is the classification (\ref{Red})
and the possible reductions of the orbital algebra $U_L(6)$ as in IBM-1:
\begin{equation}
U(18)\supset
\left( U_L(6)\supset
\left\{\begin{array}{c}
U_L(5)\supset O_L(5)\\
SU_L(3)\\
O_L(6)\supset O_L(5)
\end{array}\right\}
\supset O_L(3)
\right)
\otimes \Big( SU_T(3)\supset O_T(3)
\Big).
\label{ChainT}
\end{equation}
The most general IBM-3 Hamiltonian
{\em that conserves $U_L(6)$ or $SU_T(3)$ symmetry}
consists of a combination of Casimir operators of the algebras
appearing in (\ref{ChainT}).
If up to two-body terms are considered,
the following Casimir form results:
\begin{eqnarray}
\hat{H}_{\rm cas}&=&
A_1\hat{C}_1[U_L(6)]+
A_2\hat{C}_2[U_L(6)]+
B_1\hat{C}_1[U_L(5)]+
B_2\hat{C}_2[U_L(5)]
\nonumber\\&+&
C_2\hat{C}_2[SU_L(3)]+
D_2\hat{C}_2[O_L(6)]+
E_2\hat{C}_2[O_L(5)]+
F_2\hat{C}_2[O_L(3)]
\nonumber\\&+&
\alpha_2\hat{C}_2[SU_T(3)]+
\beta_2 \hat{C}_2[O_T(3)],
\label{HamCas}
\end{eqnarray}
where $\hat{C}_n[G]$
denotes the $n$th order Casimir operator of the algebra $G$.
No terms in $\hat{C}_1[O_L(2)]$ or $\hat{C}_1[O_T(2)]$ are added
since they would break $O_L(3)$ or $O_T(3)$,
which are assumed to be true symmetries of the Hamiltonian.
It must be emphasized that $\hat{H}_{\rm cas}$
corresponds to a subclass
of the general Hamiltonians (\ref{HamSta}) or (\ref{HamMul})
and this is so since all Casimir operators in (\ref{HamCas})
belong to the reduction scheme (\ref{ChainT}).
As already mentioned, alternative classifications exist
that conserve $L$ and $T$,
and Casimir operators of those alternative algebras would be required
to construct the most general IBM-3 Hamiltonian.
The criterion for the choice of (\ref{HamCas})
is simply that it consists of all linear and quadratic Casimir operators
of subalgebras appearing in (\ref{ChainT}).

The relation between the Hamiltonians in standard and Casimir form
cannot be expressed in a compact way
and it is more convenient to write the equations
that should be satisfied by the standard parameters
in the Hamiltonian (\ref{HamSta})
in order that it reduces to the form (\ref{HamCas}).
These relations read
\begin{eqnarray}
\begin{array}{l}
v_{0022}^{02}=v_{0022}^{00}~,~v_{0222}^{22}=v_{0222}^{22},\\
v_{0000}^{02}-v_{0000}^{00}=v_{0202}^{22}-v_{0202}^{20}=
v_{2222}^{02}-v_{2222}^{00}
=v_{2222}^{22}-v_{2222}^{20}=v_{2222}^{42}-v_{2222}^{40},\\
v_{0000}^{00}=v_{0202}^{20}-{1\over \sqrt{5}} v_{0022}^{00},
v_{0000}^{02}=v_{0202}^{22}-{1\over \sqrt{5}} v_{0022}^{02},\\
v_{2222}^{11}=-{8\over 7} v_{2222}^{40}+{1\over 7} v_{2222}^{20}
-{3\over \sqrt{14}} v_{0222}^{20}+v_{0202}^{20}+v_{0202}^{21},\\
v_{2222}^{31}=-{3\over 7} v_{2222}^{40}-{4\over 7} v_{2222}^{20}
+\sqrt{2 \over 7} v_{0222}^{20}+v_{0202}^{20}+v_{0202}^{21}.
\end{array}
\end{eqnarray}

A frequently used term in the Hamiltonian
is the so-called Majorana interaction
which is a pure two-body operator
that gives zero acting on states with full symmetry $[N]$ in $U_L(6)$
and non-zero on non-symmetric states.
Such an operator has the form
\begin{eqnarray}
\hat{M}&=&
-\sqrt{15}
\Big((s^\dag\times d^\dag)^{(2,1)}\times
 (\tilde{s}\times\tilde{d})^{(2,1)}\Big)^{(0,0)}_{0,0}
\nonumber\\
&+&{1\over 2}\sum_{L=1,3}\sqrt{3(2L+1)}
\Big((d^\dag\times d^\dag)^{(L,1)}\times
 (\tilde{d}\times\tilde{d})^{(L,1)}\Big)^{(0,0)}_{0,0},
\label{Maj1}
\end{eqnarray}
and can be expressed
in terms of the linear and quadratic operators of $U_L(6)$,
\begin{equation}
\hat{M}=\frac{1}{4}\Big(N(N+5)-\hat{C}_2[U_L(6)]\Big).
\label{Maj2}
\end{equation}
As already mentioned,
the existence of a sizeable Majorana interaction in the Hamiltonian
is essential for an approximate decoupling
between orbital and isospin spaces
in cases when $U_L(6)$ symmetry is not exactly conserved.

The eigenvalue problem associated
with the Hamiltonians (\ref{HamSta}), (\ref{HamMul}), or (\ref{HamCas})
should, in general, be solved numerically.
Computer codes exist \cite{Isacunp,Evan98}
that diagonalize the IBM-3 Hamiltonian
given in any of the three forms
and subsequently calculate electromagnetic transition probabilities.
Some special Hamiltonians can be solved analytically
and these are considered in the next section. 

\section{Dynamical Symmetries}
\label{sec-dyn-sym}
We start by discussing some symmetry properties
of the Hamiltonian (\ref{HamCas})
that are valid for any combination of parameters.
In general, an eigenstate of (\ref{HamCas}) can be written as
\begin{equation}
|[N_1,N_2,N_3]\phi LM_L;TM_T\rangle.
\label{Sta}
\end{equation}
The labels $[N_1,N_2,N_3]$, $L$, $M_L$, $T$, and $M_T$
are always good quantum numbers
because the algebras $U_L(6)$, $O_L(3)$, $O_L(2)$, $O_T(3)$, and $O_T(2)$
are common to all classifications in (\ref{ChainT}).
The $(\lambda_T,\mu_T)$ of $SU_T(3)$ are not shown in (\ref{Sta})
since they are equivalent to $[N_1,N_2,N_3]$;
all additional labels necessary to completely specify the state
are denoted by $\phi$.

A related consequence of the conservation of quantum numbers is that
the associated Casimir operators are diagonal in the basis (\ref{Sta}):
\begin{eqnarray}
\langle\hat{C}_1[U_L(6)]\rangle&=&
N_1+N_2+N_3,
\nonumber\\
\langle\hat{C}_2[U_L(6)]\rangle&=&
N_1(N_1+5)+N_2(N_2+3)+N_3(N_3+1),
\nonumber\\
\langle\hat{C}_2[O_L(3)]\rangle&=&
L(L+1),
\nonumber\\
\langle\hat{C}_2[SU_T(3)]\rangle&=&
\lambda_T^2+\mu_T^2+3(\lambda_T+\mu_T)+\lambda_T\mu_T,
\nonumber\\
\langle\hat{C}_2[O_T(3)]\rangle&=&
T(T+1),
\label{DiaMat}
\end{eqnarray}
where $\langle\hat{C}_n[G]\rangle$ denotes
$\langle[N_1,N_2,N_3]\phi LM_L;TM_T
|\hat{C}_n[G]
|[N_1,N_2,N_3]\phi LM_L;TM_T\rangle$.
In the symmetric representation $[N]$ of $U(18)$
the Casimir operators $\hat{C}_2[U_L(6)]$ and $\hat{C}_2[SU_T(3)]$
are not independent but they are related through
\begin{equation}
\hat{C}_2[SU_T(3)]=
{3\over2}\hat{C}_2[U_L(6)]-{9\over 2}N-{1\over2}N^2.
\label{U6SU3}
\end{equation} 

The isospin $T$ values contained in a given $U_L(6)$ representation
can be obtained from the reduction
of the corresponding $SU_T(3)$ representation to $O_T(3)$.
The $SU(3)\supset O(3)$ branching rule
is known in general from \cite{Elli58};
for the lowest $U_L(6)$ representations one finds
\begin{eqnarray}
[N]     &:& T=N,N-2,\dots,1\;{\rm or}\;0,
\nonumber\\[0mm]
[N-1,1] &:& T=N-1,N-2,\dots,1.
\label{t-content}
\end{eqnarray}

\subsection{The $U(5)$ Limit}
\label{sec-u5-lim}
The orbital reduction in this limit is
\begin{equation}
\begin{array}{ccccccccc}
U_L(6)&\supset&U_L(5)&\supset&O_L(5)&\supset&O_L(3)&\supset&O_L(2)\\
\downarrow&&\downarrow&&\downarrow&&\downarrow&&\downarrow\\
{[N_1,N_2,N_3]}&&(n_1,n_2,n_3)&&(\upsilon_1,\upsilon_2)&\alpha& L&&M_L
\end{array},
\label{RedU5}
\end{equation}
where $\alpha$ is a missing label,
necessary to completely specify the $O(5)\supset O(3)$ reduction.
Wave functions in this limit are thus characterized by
\begin{equation}
|[N_1,N_2,N_3](n_1,n_2,n_3)(\upsilon_1,\upsilon_2)\alpha LM_L;TM_T\rangle.
\label{StaU5}
\end{equation}
The lowest $U(5)$ eigenstates are listed in Table~\ref{tab-sta-u5}
together with a short-hand notation for them.
Reduction rules for symmetric $[N]$ and non-symmetric $[N-1,1]$ states
can be found in \cite{Arim76} and \cite{Isac84}, respectively.
 
The Hamiltonian in this limit is
\begin{eqnarray}
\hat{H}_{\rm cas}&=&
A_1\hat{C}_1[U_L(6)]+
A_2\hat{C}_2[U_L(6)]+
B_1\hat{C}_1[U_L(5)]+
B_2\hat{C}_2[U_L(5)]
\nonumber\\&+&
E_2\hat{C}_2[O_L(5)]+
F_2\hat{C}_2[O_L(3)]+
\alpha_2\hat{C}_2[SU_T(3)]+
\beta_2 \hat{C}_2[O_T(3)],
\label{HamU5}
\end{eqnarray}
with eigenvalues
\begin{eqnarray}
E&=&
A_1(N_1+N_2+N_3)+
A_2[N_1(N_1+5)+N_2(N_2+3)+N_3(N_3+1)]
\nonumber\\&+&
B_1(n_1+n_2+n_3)+
B_2[n_1(n_1+4)+n_2(n_2+2)+n_3^2]
\nonumber\\&+&
E_2[\upsilon_1(\upsilon_1+3)+\upsilon_2(\upsilon_2+1)]+
F_2L(L+1)
\nonumber\\&+&
\alpha_2[\lambda_T^2+\mu_T^2+3(\lambda_T+\mu_T)+\lambda_T\mu_T]+
\beta_2 T(T+1).
\label{EigU5}
\end{eqnarray}
A typical energy spectrum is shown in Fig.~\ref{u5-spec}.

\subsection{The $SU(3)$ Limit}
\label{sec-su3-lim}
The orbital reduction in this limit is
\begin{equation}
\begin{array}{ccccccccc}
U_L(6)&\supset&SU_L(3)&\supset&O_L(3)&\supset&O_L(2)\\
\downarrow&&\downarrow&&\downarrow&&\downarrow\\
{[N_1,N_2,N_3]}&\beta&(\lambda,\mu)&\kappa& L&&M_L
\end{array},
\label{RedSU3}
\end{equation}
where $\beta$ and $\kappa$ are missing labels,
necessary to completely specify the $U(6)\supset SU(3)$
and $SU(3)\supset O(3)$ reductions.
Wave functions in this limit are thus characterized by
\begin{equation}
|[N_1,N_2,N_3]\beta(\lambda,\mu)\kappa LM_L;TM_T\rangle.
\label{StaSU3}
\end{equation}
The lowest $SU(3)$ eigenstates are listed in Table~\ref{tab-sta-su3}
together with a short-hand notation for them.
Reduction rules for symmetric $[N]$ and non-symmetric $[N-1,1]$ states
can be found in \cite{Arim78} and \cite{Isac84}, respectively.

The Hamiltonian associated with this group chain is
\begin{eqnarray}
\hat{H}_{\rm cas}&=&
A_1\hat{C}_1[U_L(6)]+
A_2\hat{C}_2[U_L(6)]+
C_2\hat{C}_2[SU_L(3)]+
F_2\hat{C}_2[O_L(3)]
\nonumber\\&+&
\alpha_2\hat{C}_2[SU_T(3)]+
\beta_2 \hat{C}_2[O_T(3)],
\label{HamSU3}
\end{eqnarray}
with eigenvalues
\begin{eqnarray}
E&=&
A_1(N_1+N_2+N_3)+
A_2[N_1(N_1+5)+N_2(N_2+3)+N_3(N_3+1)]
\nonumber\\&+&
C_2[\lambda^2+\mu^2+3(\lambda+\mu)+\lambda\mu]+
F_2L(L+1)
\nonumber\\&+&
\alpha_2[\lambda_T^2+\mu_T^2+3(\lambda_T+\mu_T)+\lambda_T\mu_T]+
\beta_2 T(T+1).
\label{EigSU3}
\end{eqnarray}
A typical energy spectrum is shown in Fig.~\ref{su3-spec}.

\subsection{The $O(6)$ Limit}
\label{sec-o6-lim}
The orbital reduction in this limit is
\begin{equation}
\begin{array}{ccccccccc}
U_L(6)&\supset&O_L(6)&\supset&O_L(5)&\supset&O_L(3)&\supset&O_L(2)\\
\downarrow&&\downarrow&&\downarrow&&\downarrow&&\downarrow\\
{[N_1,N_2,N_3]}&&(\sigma_1,\sigma_2,\sigma_3)
&&(\upsilon_1,\upsilon_2)&\alpha& L&&M_L
\end{array},
\label{RedO6}
\end{equation}
where $\alpha$ is a missing label,
necessary to completely specify the $O(5)\supset O(3)$ reduction.
Wave functions in this limit are thus characterized by
\begin{equation}
|[N_1,N_2,N_3](\sigma_1,\sigma_2,\sigma_3)
(\upsilon_1,\upsilon_2)\alpha LM_L;
TM_T\rangle.
\label{StaO6}
\end{equation}
The lowest $O(6)$ eigenstates are listed in Table~\ref{tab-sta-o6}
together with a short-hand notation for them.
Reduction rules for symmetric $[N]$ and non-symmetric $[N-1,1]$ states
can be found in \cite{Arim79} and \cite{Isac84}, respectively.

The Hamiltonian associated with this group chain is
\begin{eqnarray}
\hat{H}_{\rm cas}&=&
A_1\hat{C}_1[U_L(6)]+
A_2\hat{C}_2[U_L(6)]+
D_2\hat{C}_2[O_L(6)]+
E_2\hat{C}_2[O_L(5)]
\nonumber\\&+&
F_2\hat{C}_2[O_L(3)]+
\alpha_2\hat{C}_2[SU_T(3)]+
\beta_2 \hat{C}_2[O_T(3)],
\label{HamO6}
\end{eqnarray}
with eigenvalues
\begin{eqnarray}
E&=&
A_1(N_1+N_2+N_3)+
A_2[N_1(N_1+5)+N_2(N_2+3)+N_3(N_3+1)]
\nonumber\\&+&
D_2[\sigma_1(\sigma_1+4)+\sigma_2(\sigma_2+2)+\sigma_3^2]
\nonumber\\&+&
E_2[\upsilon_1(\upsilon_1+3)+\upsilon_2(\upsilon_2+1)]+
F_2L(L+1)
\nonumber\\&+&
\alpha_2[\lambda_T^2+\mu_T^2+3(\lambda_T+\mu_T)+\lambda_T\mu_T]+
\beta_2 T(T+1).
\label{EigO6}
\end{eqnarray}
A typical energy spectrum is shown in Fig.~\ref{o6-spec}.

\section{Electromagnetic Transitions}
\label{sec-elec-tran}
A general one-body electromagnetic operator in IBM-3
consists of  isoscalar, isovector, and isotensor parts,
\begin{equation}
\hat{T}^{(l)}_{m_l}(l_1,l_2)=
a_0\hat{T}^{(l,0)}_{m_l,0}(l_1,l_2)+
a_1\hat{T}^{(l,1)}_{m_l,0}(l_1,l_2)+
a_2\hat{T}^{(l,2)}_{m_l,0}(l_1,l_2),
\label{Ope}
\end{equation}
where, as in (\ref{OpeMul}), the superscripts in the operators
on the rhs refer to the angular momentum and the isospin, respectively,
whereas on the lhs only the angular momentum is given
since the operator corresponds to an admixture of isospins.
In previous IBM-3 studies only isoscalar and isovector 
electromagnetic operators are considered \cite{Elli96-2};
isotensor contributions are included in this paper for completeness.
The relevance of this contribution
is discussed in Section~\ref{sec-exp}
when comparing with experimental results.
The parameters $a_t$ are boson $g$ factors, boson electric charges, etc.
depending on the multipolarity of the operator,
and the operators $\hat{T}^{(l,t)}_{m_l,0}(l_1,l_2)$ are defined as
\begin{eqnarray}
\hat{T}^{(l,0)}_{m_l,0}(l_1,l_2)&=&
\sqrt{3}
(b^\dag_{l_1,1}\times\tilde{b}_{l_2,1})^{(l,0)}_{m_l,0},
\nonumber\\
\hat{T}^{(l,1)}_{m_l,0}(l_1,l_2)&=&
\sqrt{2}
(b^\dag_{l_1,1}\times\tilde{b}_{l_2,1})^{(l,1)}_{m_l,0},
\nonumber\\
\hat{T}^{(l,2)}_{m_l,m_t}(l_1,l_2)&=&
-\sqrt{6}
(b^\dag_{l_1,1}\times\tilde{b}_{l_2,1})^{(l,2)}_{m_l,0}.
\label{OpeDef}
\end{eqnarray}
The factors are taken for later convenience 
and lead to the explicit forms
\begin{eqnarray}
\hat{T}^{(l,0)}_{m_l,0}(l_1,l_2)&=&
(b^\dag_{l_1,\pi}\times\tilde{b}_{l_2,\pi})^{(l)}_{m_l}+
(b^\dag_{l_1,\delta}\times\tilde{b}_{l_2,\delta})^{(l)}_{m_l}+
(b^\dag_{l_1,\nu}\times\tilde{b}_{l_2,\nu})^{(l)}_{m_l},
\nonumber\\
\hat{T}^{(l,1)}_{m_l,0}(l_1,l_2)&=&
(b^\dag_{l_1,\pi}\times\tilde{b}_{l_2,\pi})^{(l)}_{m_l}-
(b^\dag_{l_1,\nu}\times\tilde{b}_{l_2,\nu})^{(l)}_{m_l},
\nonumber\\
\hat{T}^{(l,2)}_{m_l,0}(l_1,l_2)&=&
(b^\dag_{l_1,\pi}\times\tilde{b}_{l_2,\pi})^{(l)}_{m_l}+
2(b^\dag_{l_1,\delta}\times\tilde{b}_{l_2,\delta})^{(l)}_{m_l}-
(b^\dag_{l_1,\nu}\times\tilde{b}_{l_2,\nu})^{(l)}_{m_l}
\label{OpeIso}
\end{eqnarray}
We may thus write the operators (\ref{Ope}) alternatively as
\begin{equation}
\hat{T}^{(l)}_{m_l}(l_1,l_2)=
a_\pi(b^\dag_{l_1,\pi}\times\tilde{b}_{l_2,\pi})^{(l)}_{m_l}+
a_\delta(b^\dag_{l_1,\delta}\times\tilde{b}_{l_2,\delta})^{(l)}_{m_l}+
a_\nu(b^\dag_{l_1,\nu}\times\tilde{b}_{l_2,\nu})^{(l)}_{m_l},
\label{OpeAlt}
\end{equation}
where the $a_\rho$ with $\rho=\pi,\delta,\nu$
are related to the $a_t$ through
\begin{equation}
a_0=a_\pi+a_\delta+a_\nu,
\qquad
a_1=a_\pi-a_\nu,
\qquad
a_2=-a_\pi+2a_\delta-a_\nu.
\label{ParRel}
\end{equation}

For the calculation of matrix elements
it is also of interest to know
the tensor properties of the transition operators
under $SU_T(3)\supset O_T(3)\supset O_T(2)$.
Given that $s^\dag$ or $d^\dag_\mu$ transforms
as a $\hat{T}^{(1,0)1}$ tensor under $SU_T(3)\supset O_T(3)$
and that $s$ or $\tilde{d}_\mu$ transforms as $\hat{T}^{(0,1)1}$,
one finds that the full tensor character of the coupled operator
is uniquely determined by its isospin coupling, that is,
\begin{eqnarray}
\hat{T}^{(l,0)}_{m_l,0}(l_1,l_2)&\rightarrow&\hat{T}^{(0,0)00},
\nonumber\\
\hat{T}^{(l,1)}_{m_l,m_t}(l_1,l_2)&\rightarrow&\hat{T}^{(1,1)1m_t},
\nonumber\\
\hat{T}^{(l,2)}_{m_l,m_t}(l_1,l_2)&\rightarrow&\hat{T}^{(1,1)2m_t}.
\label{TenCha}
\end{eqnarray}
Since the isovector and isotensor operators
belong to the same $SU(3)$ representation (1,1),
their $O_T(3)\subset SU_T(3)$ reduced matrix elements are related,
\begin{equation}
{{\langle[N_1,N_2,N_3]\alpha
\|\hat T^{(l,2)}_{m_l,*}(l_1,l_2)
\|[N'_1,N'_2,N'_3]\alpha'\rangle}
\over
{\langle[N_1,N_2,N_3]\alpha
\|\hat T^{(l,1)}_{m_l,*}(l_1,l_2)
\|[N'_1,N'_2,N'_3]\alpha'\rangle}}
=-\sqrt{3}.
\label{Norma4}
\end{equation}
This ratio is different from one
due to the specific normalization of the tensor operators (\ref{OpeDef}).

The orbital structure of the electromagnetic transition operators
is the usual one as it occurs, for example, in IBM-1 or IBM-2.
In particular, for M1, E2, and M3 transitions
the following form is taken:
\begin{eqnarray}
\hat{T}({\rm M}1)&=&
\sqrt{\frac{3}{4\pi}}
g\sqrt{10}(d^\dag\times\tilde{d})^{(1)},
\nonumber\\
\hat{T}({\rm E}2)&=&
e\Big((s^\dag\times\tilde{d}+d^\dag\times\tilde{s})^{(2)}
 +\chi(d^\dag\times\tilde{d})^{(2)}\Big),
\nonumber\\
\hat{T}({\rm M}3)&=&
\sqrt{\frac{35}{8\pi}}
\Omega(d^\dag\times\tilde{d})^{(3)}.
\label{M1E2M3}
\end{eqnarray}

\subsection{Symmetric-to-Symmetric Transitions}
\label{sec-sym-tran}
Symmetric states of the IBM-3
are defined as having $U_L(6)$ quantum numbers
$[N_1,N_2,N_3]=[N,0,0]\equiv[N]$
or, equivalently, $SU_T(3)$ quantum numbers
$(\lambda_T,\mu_T)=(N,0)$.
There is a one-to-one correspondence
between the symmetric states of IBM-3 and all states of IBM-1
and, because of this, matrix elements between symmetric IBM-3 states
can be related to the corresponding ones in IBM-1,
as shown in this subsection.

The starting point is to consider
a matrix element between IBM-3 states with $M_T=-N$ (all-neutron states),
for which, by virtue of (\ref{OpeIso}),
the following relations are satisfied: 
\begin{eqnarray}
\langle[N]\phi'L';N,-N
\|\hat{T}^{(l,0)}_{*,0}(l_1,l_2)                                 
\|[N]\phi L;N-N\rangle_{\rm IBM3}
&=&
\langle[N]\phi'L'
\|\hat{T}^{(l)}(l_1,l_2)
\|[N]\phi L\rangle_{\rm IBM1},
\nonumber\\
\langle[N]\phi'L';N,-N
\|\hat{T}^{(l,1)}_{*,0}(l_1,l_2)
\|[N]\phi L;N,-N\rangle_{\rm IBM3}
&=&
-\langle[N]\phi'L'
\|\hat{T}^{(l)}(l_1,l_2)
\|[N]\phi L\rangle_{\rm IBM1},
\nonumber\\
\langle[N]\phi'L';N,-N
\|\hat{T}^{(l,2)}_{*,0}(l_1,l_2)
\|[N]\phi L;N,-N\rangle_{\rm IBM3}
&=&
-\langle[N]\phi'L'
\|\hat{T}^{(l)}(l_1,l_2)
\|[N]\phi L\rangle_{\rm IBM1},
\end{eqnarray}
where $\phi$ represents any label in between $U_L(6)$ and $O_L(3)$
and all states have $T=-M_T=N$.
The symbols $\|$ indicate a matrix element
reduced from $O_L(2)$ to $O_L(3)$;
no $O_L(2)$ indices are required in bra, ket, or operator,
and in the latter this is indicated by an asterisk.
To find the general relationship
between corresponding IBM-1 and IBM-3 matrix elements,
one may use the Wigner--Eckart theorem for $O_T(3)\subset SU_T(3)$.
Expressions can be derived for any $T$ and $M_T$,
but only the case $T=-M_T$ is listed here
since it corresponds to states lowest in energy.
(An exception to this rule occurs in self-conjugate odd--odd nuclei
where $T=0$ and $T=1$ states are close in energy;
these nuclei, however, are not amenable to the IBM-3 description
considered in this paper.)
For an isoscalar operator one finds the relation
\begin{eqnarray}
\lefteqn{
{{\langle[N]\phi'L';T,-T
\|\hat{T}^{(l,0)}_{*,0}(l_1,l_2)
\|[N]\phi L;T,-T\rangle_{\rm IBM3}}
\over     
{\langle[N]\phi'L';N,-N
\|\hat{T}^{(l,0)}_{*,0}(l_1,l_2)
\|[N]\phi L;N,-N\rangle_{\rm IBM3}}}}
\nonumber\\
&=&
{{\left\langle\begin{array}{cc|c}
(N,0)&(0,0)&(N,0)\\
  T  &  0  &  T
\end{array}\right\rangle
\langle T-T\;00|T-T\rangle}
\over
{\left\langle\begin{array}{cc|c}
(N,0)&(0,0)&(N,0)\\
  N  &  0  &  N
\end{array}\right\rangle
\langle N-N\;00|N-N\rangle}},
\label{isosc1}
\end{eqnarray}
where the symbols between big angle brackets
are $SU(3)\supset O(3)$ isoscalar factors \cite{Verg68},
which in this case trivially are equal to one.
The following relation is thus found for an isoscalar operator:
\begin{equation}
\langle[N]\phi'L';T,-T
\|\hat{T}^{(l,0)}_{*,0}(l_1,l_2)
\|[N]\phi L;T,-T\rangle_{\rm IBM3}
=
\langle[N]\phi'L'
\|\hat{T}^{(l)}(l_1,l_2)
\|[N]\phi L\rangle_{\rm IBM1}.
\label{isosc2}
\end{equation}
The analogous relation for isovector ($t=1$) and isotensor ($t=2$) operators is
\begin{eqnarray}
\lefteqn{
{{\langle[N]\phi'L';T,-T
\|\hat{T}^{(l,t)}_{*,0}(l_1,l_2)
\|[N]\phi L ;T,-T\rangle_{\rm IBM3}}
\over
{\langle[N]\phi'L';N,-N
\|\hat{T}^{(l,t)}_{*,0}(l_1,l_2)
\|[N]\phi L;N,-N\rangle_{\rm IBM3}}}}
\nonumber\\
&=&
{{\left\langle\begin{array}{cc|c}
(N,0)&(1,1)&(N,0)\\
  T  &  t  &  T
\end{array}\right\rangle
\langle T-T\;t0|T-T\rangle}
\over
{\left\langle\begin{array}{cc|c}
(N,0)&(1,1)&(N,0)\\
  N  &  t  &  N
\end{array}\right\rangle
\langle N-N\;t0|N-N\rangle}},
\label{iso-vt1}
\end{eqnarray}
such that
\begin{eqnarray}
\lefteqn{
\langle[N]\phi'L';T,-T
\|\hat{T}^{(l,1)}_{*,0}(l_1,l_2)
\|[N]\phi L;T,-T\rangle_{\rm IBM3}}
\nonumber\\
&=&
-{{T}\over{N}}
\langle[N]\phi'L'
\|\hat{T}^{(l)}(l_1,l_2)
\|[N]\phi L\rangle_{\rm IBM1},
\nonumber\\
\lefteqn{
\langle[N]\phi'L';T,-T
\|\hat{T}^{(l,2)}_{*,0}(l_1,l_2)
\|[N]\phi L;T,-T\rangle_{\rm IBM3}}
\nonumber\\
&=&
-{{T(2N+3)}\over{(2T+3)N}}
\langle[N]\phi'L'
\|\hat{T}^{(l)}(l_1,l_2)
\|[N]\phi L\rangle_{\rm IBM1},
\nonumber\\
\lefteqn{
\langle[N]\phi'L';T+2,-T
\|\hat{T}^{(l,2)}_{*,0}(l_1,l_2)
\|[N]\phi L;T,-T\rangle_{\rm IBM3}}
\nonumber\\
&=&
6\sqrt{{(T+1)(N-T)(N+T+3)}\over{(2T+3)^2(2T+5)N^2}}
\langle[N]\phi'L'
\|\hat{T}^{(l)}(l_1,l_2)
\|[N]\phi L\rangle_{\rm IBM1}.
\label{iso-scvt1}
\end{eqnarray}
Note that all results are symmetric
under interchange of the orbital parts $\phi L$ and $\phi' L'$.
An alternative derivation of these results
can be found in \cite{Abde88}.

Expressions for symmetric-to-symmetric transitions in IBM-3
can now be derived from the corresponding ones in IBM-1
\cite{Arim76,Arim78,Arim79}.
In Tables~\ref{tab-tran-u5}, \ref{tab-tran-su3-1}, \ref{tab-tran-su3-2},
and \ref{tab-tran-o6}
all non-zero M1, E2, and M3 transitions out of the ground state
are listed for the U(5), SU(3), and O(6) limits.
Isoscalar, isovector, and isotensor parts are given separately;
the E2 operator is defined with $\chi=-\frac{\sqrt{7}}{2}$ and $\chi=0$
in the $SU(3)$ and $O(6)$ limits, respectively.

\subsection{Symmetric-to-Non-Symmetric Transitions}
\label{sec-nonsym-tran}
Generally, non-symmetric states in IBM-3
have $U_L(6)$ quantum numbers different from $[N]$.
Usually, however, only $[N-1,1]$ states are considered in the analysis
and this is what will be done here.
Just as there exists a one-to-one correspondence
between IBM-1 and symmetric IBM-3 states,
a one-to-one correspondence can be established
between IBM-2 and non-symmetric, two-rowed IBM-3 states.
This fact can be exploited to derive relations
between matrix elements involving non-symmetric $[N-1,1]$ states in IBM-3
and corresponding ones in IBM-2 known from \cite{Isac86}.

The IBM-2 states
that can be related to the IBM-3 classification (\ref{ChainT})
are classified themselves according to
\begin{eqnarray}
\begin{array}{l}
U(12)\supset
\left( U_L(6)\supset
\left\{\begin{array}{c}
U_L(5)\supset O_L(5)\\
SU_L(3)\\
O_L(6)\supset O_L(5)
\end{array}\right\}
\supset O_L(3)
\right)
\otimes SU_F(2).
\end{array}
\label{ChainF}
\end{eqnarray}
Comparison of (\ref{ChainT}) and (\ref{ChainF})
shows that the correspondence between IBM-2 and IBM-3
can be established via the algebra $U_L(6)$.
All IBM-2 states belong to $U_L(6)$ representations of the type $[N-f,f]$
where $f=0,1,\dots,\min(N_\pi,N_\nu)$
and these form a subset of the possible $U_L(6)$ representations in IBM-3.
The $F$ spin is defined as $F=N/2-f$.
An $F$-scalar operator cannot contribute
to symmetric-to-non-symmetric transitions.
Since a one-body operator is either $F$ scalar or $F$ vector,
only the latter can connect symmetric with a non-symmetric state.
Furthermore, an $F$-vector operator in IBM-2
coincides with its isovector counterpart in IBM-3
[see (\ref{ParRel})],
\begin{equation}
\hat T^{(L,F=1)}_{M_L,M_F} (l_1,l_2)=
\hat T^{(L,T=1)}_{M_L,M_T} (l_1,l_2).
\label{Fvect}
\end{equation}

To obtain the relation between IBM-2 and IBM-3 transitions,
one must first establish the tensor character
of the transition operator under $U_L(6)$.
The boson operators $b^\dag_{lm,1\mu}$
transform as a $\hat T^{[1]}$ tensor operator under $U_L(6)$
while $\tilde b_{lm,1\mu}$ transforms as $\hat T^{[1^5]}$.
Consequently, the $U_L(6)$ tensor character
of an $F$-vector or isovector one-body operator is $\hat T^{[2,1^4]}$.
Because of this property
the ratio between corresponding $[N]\rightarrow[N-1,1]$ matrix elements
of an $F$-vector operator in IBM-2
and an isovector operator in IBM-3,
is a unique function of $N_\pi$, $N_\nu$, $N$, and $T$,
independent of the particular states or of the operator.
For simplicity the $SU_L(3)$ limit is analyzed
but the proof can be extended to any limit,
the only requirement being that it has $U_L(6)$ symmetry.
First the ratio of arbitrary matrix elements
can be shown to be related to a ratio of specific ones,
\begin{eqnarray}
\lefteqn{
{{\langle[N-1,1]\beta'(\lambda',\mu')\kappa'L';T,-T
\|\hat T^{[2,1^4](\bar{\lambda},\bar{\mu})(l,1)}_{*,0}
\|[N]\beta(\lambda,\mu)\kappa L;T,-T\rangle_{\rm IBM3}}
\over
{\langle[N-1,1]\beta'(\lambda',\mu')\kappa'L'
\|\hat T^{[2,1^4](\bar{\lambda},\bar{\mu})(l,1)}_{*,0}
\|[N]\beta(\lambda,\mu)\kappa L\rangle_{\rm IBM2}}}}
\nonumber\\
&=&
{{\langle[N-1,1]1_M^+;T,-T
\|\hat T(M1)
\|[N]0_1^+;T,-T\rangle_{\rm IBM3}}
\over
{\langle[N-1,1]1_M^+ 
\|\hat T(M1)
\|[N]0_1^+\rangle_{\rm IBM2}}},
\label{Ratio1}
\end{eqnarray}
where $\|$ indicates a $O_L(2)\subset O_L(3)$ reduced matrix element.
The identity (\ref{Ratio1})
follows from the Wigner--Eckart theorem
in $O_L(3)\subset SU_L(3)\subset U_L(6)$,
which leads to the same $U_L(6)$-reduced matrix element
in both the general and the specific case.
Hence
\begin{eqnarray}
\lefteqn{
\langle[N-1,1]\beta'(\lambda',\mu')\kappa'L'
\|\hat T^{[2,1^4](\bar{\lambda},\bar{\mu})(l,1)}_{*,0}
\|[N]\beta(\lambda,\mu)\kappa L\rangle}
\nonumber\\
&=&
\langle[N-1,1]1_M^+
\|\hat T(M1)
\|[N]0_1^+\rangle
\nonumber\\ 
&\times&
{{\left\langle\begin{array}{cc|c}
[N]                 & [2,1^4]                   & [N-1,1]\\
\beta (\lambda,\mu) & (\bar{\lambda},\bar{\mu}) & \beta' (\lambda',\mu')
\end{array}\right\rangle
\left\langle\begin{array}{cc|c}
(\lambda,\mu)    & (\bar{\lambda},\bar{\mu}) & (\lambda',\mu')\\
\kappa L         & l                         & \kappa' L'
\end{array}\right\rangle}
\over
{\left\langle\begin{array}{cc|c} 
[N]    & [2,1^4] & [N-1,1]\\
(2N,0) & (1,1)   & (2N-2,1)
\end{array}\right\rangle
\left\langle\begin{array}{cc|c}
(2N,0) & (1,1) & (2N-2,1)\\
0      & 1     & 1
\end{array}\right\rangle}},
\label{Ratio2}
\end{eqnarray}
where the symbols between big angle brackets
are $U(6)\supset SU(3)$ or $SU(3)\supset O(3)$ isoscalar factors.
The $F$-spin and isospin labels can be omitted from the matrix elements
because the result (\ref{Ratio2}) is identical in IBM-2 and IBM-3.
Taking the ratio of an IBM-2 and an IBM-3 matrix element,
the isoscalar factors cancel out and the result (\ref{Ratio1})
is obtained.
Since both matrix elements on the rhs of (\ref{Ratio1})
are known from \cite{Isac86} and \cite{Hals95},
and since the derivation can be generalized to any state
with good $U_L(6)$ symmetry,
the following relation results:
\begin{eqnarray}
\lefteqn{
\langle[N-1,1]\phi'L';T,-T
\|\hat T^{[2,1^4](l,1)}_{*,0}
\|[N]\phi L;T,-T\rangle_{\rm IBM3}}
\nonumber\\
&=&
\sqrt{{1\over{4N_\pi N_\nu}}{{T(N-T)(N+T+1)}\over{T+1}}}
\langle[N-1,1]\phi'L'
\|\hat T^{[2,1^4](l,1)}_{*,0}
\|[N]\phi L\rangle_{\rm IBM2}.
\label{Ratio3}
\end{eqnarray}
It is clear that this ratio
is independent of the orbital structure of the states
and/or the electromagnetic operator.
Analogous relations for isotensor operators and $T\neq |M_T|$ states
can be obtained using (\ref{Ratio3})
together with the Wigner--Eckart theorem,
\begin{eqnarray}
\lefteqn{
\langle[N-1,1]\phi'L';T+1,-T
\|\hat T^{[2,1^4](l,1)}_{*,0}
\|[N]\phi L;T,-T\rangle_{\rm IBM3}}
\nonumber\\
&=&
\sqrt{{1\over{4N_\pi N_\nu}}{{(T+2)N(N-T)}\over{(T+1)(2T+3)}}}
\langle[N-1,1]\phi'L'
\|\hat T^{[2,1^4](l,1)}_{*,0} 
\|[N]\phi L\rangle_{\rm IBM2},
\nonumber\\
\lefteqn{
\langle[N-1,1]\phi'L';T,-T
\|\hat T^{[2,1^4](l,2)}_{*,0}
\|[N]\phi L;T,-T\rangle_{\rm IBM3}}
\nonumber\\
&=&
-\sqrt{{3\over{4N_\pi N_\nu}}{{3T(N-T)(N+T+1)}\over{(T+1)(2T+3)^2}}}
\langle[N-1,1]\phi'L'
\|\hat T^{[2,1^4](l,1)}_{*,0}
\|[N]\phi L\rangle_{\rm IBM2},
\nonumber\\
\lefteqn{
\langle[N-1,1]\phi'L';T+1,-T
\|\hat T^{[2,1^4](l,2)}_{*,0}
\|[N]\phi L;T,-T\rangle_{\rm IBM3}}
\nonumber\\
&=&
\sqrt{{3\over{4N_\pi N_\nu}}{{3T^2N(N-T)}\over{(T+1)(T+2)(2T+3)}}}
\langle[N-1,1]\phi'L'
\|\hat T^{[2,1^4](l,1)}_{*,0}
\|[N]\phi L\rangle_{\rm IBM2},
\nonumber\\
\lefteqn{
\langle[N-1,1]\phi'L';T+2,-T
\|\hat T^{[2,1^4](l,2)}_{*,0}
\|[N]\phi L;T,-T\rangle_{\rm IBM3}}
\nonumber\\
&=&
\sqrt{{3\over{4N_\pi N_\nu}}
      {{12(T+1)(T+3)(N-T)(N-T-2)}\over{(T+2)(2T+3)^2(2T+5)}}}
\nonumber\\
&\times&
\langle[N-1,1]\phi'L'
\|\hat T^{[2,1^4](l,1)}_{*,0}
\|[N]\phi L\rangle_{\rm IBM2}.
\label{Ratio4}
\end{eqnarray}
Note an additional factor $-\sqrt{3}$
for the isotensor matrix elements
due to the specific normalization of the isovector and isotensor operators. 
An alternative derivation of these results can be found in \cite{Abde88}.

Expressions for symmetric-to-non-symmetric transitions in IBM-3
can now be derived from the corresponding ones in IBM-2 \cite{Isac86}.
In Tables~\ref{tab-tran-u5}, \ref{tab-tran-su3-1}, \ref{tab-tran-su3-2},
and \ref{tab-tran-o6}
all non-zero M1, E2, and M3 transitions out of the ground state
are listed for the U(5), SU(3), and O(6) limits.

\section{Two-Nucleon Transfer Probabilities}
\label{sec-two-nucl}
Two-nucleon transfer reactions have been studied in IBM-1 \cite{Iach87},
generally showing a good agreement
with experimental observations \cite{Cize81}.
To see whether any specific effects
are found for two-nucleon transfer properties
if a proton--neutron pair is included in the boson basis,
it is of interest to compare the IBM-1 analysis
with the corresponding one in IBM-3,
which is the object the present section.

A general one-boson transfer operator has the form
\begin{equation}
P^{(l)}_{+,\rho,m}=p_{\rho,l}b^\dag_{lm,1\rho},
\qquad
P^{(l)}_{-,\rho,m}=p_{\rho,l}b_{lm,1\rho},
\label{Trans1}
\end{equation}
for the addition or the removal of two nucleons, respectively,
and where $\rho=\pi,\delta,\nu$.
A measurable quantity is the transfer intensity $I$
which is defined as 
\begin{eqnarray}
I(N\phi L;\xi\rightarrow N+1\;\phi'L';\xi')
&=&
{{p_{\rho,l}^2}\over{2L+1}}
|\langle N+1\;\phi'L';\xi'
\|P^{(l)}_{+,\rho}
\|N\phi L;\xi\rangle|^2,
\nonumber\\ 
I(N\phi L;\xi\rightarrow N-1\;\phi'L';\xi')
&=&
{{p_{\rho,l}^2}\over{2L+1}}
|\langle N-1\;\phi'L';\xi'
\|P^{(l)}_{-,\rho}
\|N\phi L;\xi\rangle|^2,
\label{Trans2}
\end{eqnarray}
where $\phi$ and $\xi$
denote all additional orbital and isospin labels, respectively,
to completely specify the state
and the symbol $\|$ indicates the reduction of the matrix element
from $O_L(2)$ to $O_L(3)$.

The analysis in this section will be confined to transitions
between symmetric states with $T=-M_T$. 
Analytical expressions for transfer intensities in IBM-3
can be obtained in much the same way as in Section~\ref{sec-sym-tran}
by relating them to IBM-1
using the property that $b^\dag$ and $\tilde b$
transform as $\hat{T}^{(1,0)1}$ and $\hat{T}^{(0,1)1}$ tensors
under $SU_T(3)\supset O_T(3)$.
The starting point is again
the matrix element between IBM-3 states with $M_T=-N$,
\begin{equation}
\langle[N+1]\phi'L';N+1,-N-1
\|b^\dag_\nu
\|[N]\phi L;N,-N\rangle_{\rm IBM3}
=
\langle[N+1]\phi'L'
\|b^\dag
\|[N]\phi L\rangle_{\rm IBM1},
\label{b-neutron0}
\end{equation}
from where the following relation results:
\begin{eqnarray}
\lefteqn{
\langle[N+1]\phi'L';T-\mu,-T+\mu
\|b^\dag_{l,1\mu}
\|[N]\phi L;T,-T\rangle_{\rm IBM3}}
\nonumber\\
&=&
{{\left\langle\begin{array}{cc|c}
(N,0)&(1,0)&(N+1,0)\\
  T  & \mu & T-\mu
\end{array}\right\rangle
\langle T-T\;1\mu|T-\mu\;-T+\mu\rangle}
\over
{\left\langle\begin{array}{cc|c}
(N,0)&(1,0)&(N+1,0)\\
  N  &  1  &  N+1
\end{array}\right\rangle
\langle N-N\;1-1|N+1\;-N-1\rangle}}
\langle[N]\phi'L'
\|b^\dag_l
\|[N]\phi L\rangle_{\rm IBM1},
\label{b-all}
\end{eqnarray}
for the $\pi$ ($\mu=+1$), $\delta$ ($\mu=0$), or $\nu$ ($\mu=-1$) boson.
Insertion of the appropriate isoscalar factors and Clebsch--Gordan coefficients
leads to
\begin{eqnarray}
\lefteqn{
I([N]\phi L;T,-T\rightarrow[N+1]\phi'L';T-1,-T+1)}
\nonumber\\
&=&
{{p_{\pi,l}^2}\over{2L+1}}
{{T(N-T+2)}\over{(2T+1)(N+1)}}
\left|\langle[N+1]\phi'L'\|b^\dag_l\|[N]\phi L\rangle_{\rm IBM1}\right|^2,
\nonumber\\
\lefteqn{
I([N]\phi L;T,-T\rightarrow[N+1]\phi'L';T,-T)=0,}
\nonumber\\
\lefteqn{
I([N]\phi L;T,-T\rightarrow[N+1]\phi'L';T+1,-T-1)}
\nonumber\\
&=&
{{p_{\nu,l}^2}\over{2L+1}}
{{(T+1)(N+T+3)}\over{(2T+3)(N+1)}}
\left|\langle[N+1]\phi'L'\|b^\dag_l\|[N]\phi L\rangle_{\rm IBM1}\right|^2.
\label{Trans4}
\end{eqnarray}
From the relation
\begin{equation}
\left|\langle[N-1]\phi'L'\|\tilde b_l\|[N]\phi L\rangle_{\rm IBM1}\right|^2
=
\left|\langle[N]\phi L\|b^\dag_l\|[N-1]\phi'L'\rangle_{\rm IBM1}\right|^2,
\label{Trans7}
\end{equation}
expressions for the inverse reaction are obtained,
\begin{eqnarray}
\lefteqn{
I([N]\phi L;T,-T \rightarrow[N-1]\phi'L';T+1,-T-1)}
\nonumber\\
&=&
{{p_{\pi,l}^2}\over{2L+1}}
{{(T+1)(N-T)}\over{(2T+3)N}}
\left|\langle[N]\phi L\|b^\dag_l\|[N-1]\phi'L'\rangle_{\rm IBM1}\right|^2,
\nonumber\\
\lefteqn{
I([N]\phi L;T,-T\rightarrow[N-1]\phi'L';T,-T)=0,}
\nonumber\\
\lefteqn{
I([N]\phi L;T,-T\rightarrow[N-1]\phi'L';T-1,-T+1)}
\nonumber\\
&=&
{{p_{\nu,l}^2}\over{2L+1}}
{{T(N+T+1)}\over{(2T+1)N}}
\left|\langle[N]\phi L\|b^\dag_l\|[N-1]\phi'L'\rangle_{\rm IBM1}\right|^2.
\label{Trans8}
\end{eqnarray}
As can be seen from (\ref{Trans4}) and (\ref{Trans8})
the probability for the transfer of a $\delta$ boson is zero.
This is due to a vanishing isoscalar factor,
\begin{equation}
\left\langle\begin{array}{cc|c} 
 (N,0) & (1,0) & (N+1,0)\\
   T   &   1   &    T
\end{array}\right\rangle
=0.
\label{Isos}
\end{equation}  

To illustrate these formulas,
some particular cases in the $SU(3)$
are given in Table~\ref{tab-two-nuc}.

\section{Boson-Number Expectation Values}
\label{sec-bos-num}
The average number of bosons
is a useful quantity for comparing with other models,
for example, Shell Model Monte Carlo calculations (SMMC) \cite{Lang93}
or the isovector-pairing $SO(5)$ seniority model \cite{Enge96}.
In this section analytic expressions are derived
for the average number of the different kinds of bosons
as obtained for symmetric states with $T=|M_T|$,
though expressions for higher isospin
can be obtained through the Wigner--Eckart theorem.

Consider the following operators:
\begin{eqnarray}
\hat N^s_0
&\equiv&
(s^\dag\times\tilde{s})^{(L=0,T=0)}_{M_L=0,M_T=0}
=\frac{1}{\sqrt3}
\Big((s^\dag_\pi s_\pi)
+(s^\dag_\delta s_\delta)
+(s^\dag_\nu s_\nu)\Big),
\nonumber\\
\hat N^s_1
&\equiv&
(s^\dag\times\tilde{s})^{(L=0,T=1)}_{M_L=0,M_T=0}
=\frac{1}{\sqrt2}
\Big((s^\dag_\pi s_\pi)
-(s^\dag_\nu s_\nu)\Big),
\nonumber\\
\hat N^s_2
&\equiv&
(s^\dag\times\tilde{s})^{(L=0,T=2)}_{M_L=0,M_T=0}
=\frac{1}{\sqrt6}
\Big((s^\dag_\pi s_\pi)
-2(s^\dag_\delta s_\delta)
+(s^\dag_\nu s_\nu)\Big),
\label{S}
\end{eqnarray}
and
\begin{eqnarray}
\hat N^d_0
&\equiv&
(d^\dag\times\tilde{d})^{(L=0,T=0)}_{M_L=0,M_T=0}
=\frac{1}{\sqrt3}
\Big((d^\dag_\pi\times\tilde{d}_\pi)^{(L=0)}_{M_L=0}
+(d^\dag_\delta\times\tilde{d}_\delta)^{(L=0)}_{M_L=0}
+(d^\dag_\nu\times\tilde{d}_\nu)^{(L=0)}_{M_L=0}\Big),
\nonumber\\
\hat N^d_1
&\equiv&
(d^\dag\times\tilde{d})^{(L=0,T=1)}_{M_L=0,M_T=0}
=\frac{1}{\sqrt2}
\Big((d^\dag_\pi\times\tilde{d}_\pi)^{(L=0)}_{M_L=0}
-(d^\dag_\nu\times\tilde{d}_\nu)^{(L=0)}_{M_L=0}\Big),
\nonumber\\
\hat N^d_2
&\equiv&
(d^\dag\times\tilde{d})^{(L=0,T=2)}_{M_L=0,M_T=0}
=\frac{1}{\sqrt6}
\Big((d^\dag_\pi\times\tilde{d}_\pi)^{(L=0)}_{M_L=0}
-2(d^\dag_\delta\times\tilde{d}_\delta)^{(L=0)}_{M_L=0}
+(d^\dag_\nu\times\tilde{d}_\nu)^{(L=0)}_{M_L=0}\Big).
\label{D}
\end{eqnarray}
Boson-number operators can be defined
in terms of (\ref{S}) and (\ref{D}) by introducing
\begin{equation}
\hat N_T=\hat N^s_T+\sqrt{5}\hat N^d_T,
\label{Nt}
\end{equation}
for $T=0,1,2$.
The completely scalar operator with $L=0$ and $T=0$, $\hat N_0$,
is proportional to the number of bosons $N$.
The number operators
for the $\pi$, $\delta$, and $\nu$ bosons separately
can now be defined in terms of the $\hat N_T$,
\begin{eqnarray}
\hat N_\pi
&=&
\frac{1}{\sqrt3}\hat N_0-
\frac{1}{\sqrt2}\hat N_1-
\frac{1}{\sqrt6}\hat N_2,
\nonumber\\
\hat N_\delta
&=&
\frac{1}{\sqrt3}\hat N_0+
\sqrt{\frac{2}{3}}\hat N_2,
\nonumber\\
\hat N_{\nu}
&=&
\frac{1}{\sqrt3}\hat N_0-
\frac{1}{\sqrt2}\hat N_1+
\frac{1}{\sqrt6}\hat N_2.
\label{N}
\end{eqnarray}
For states with $M_T=-N$, $|[N]\phi L;N,-N\rangle$,
that is, for all-neutron states,
the matrix elements of $\hat N_T$ are easily evaluated,
\begin{eqnarray}
\langle[N]\phi L;N,-N|\hat N_0|[N]\phi L;N,-N\rangle
&=&-\frac{N}{\sqrt3},
\nonumber\\
\langle[N]\phi L;N,-N|\hat N_1|[N]\phi L;N,-N\rangle
&=&\frac{N}{\sqrt2},
\nonumber\\
\langle[N]\phi L;N,-N|\hat N_2|[N]\phi L;N,-N\rangle
&=&-\frac{N}{\sqrt6}.
\end{eqnarray}
To obtain expressions for the more general states $|[N]\phi L;T,-T\rangle$
use can be made of the Wigner--Eckart theorem in $O_T(3)\subset SU_T(3)$.
The results are
\begin{eqnarray}
\langle[N]\phi L;T,-T|\hat N_0|[N]\phi L;T,-T\rangle
&=&-\frac{N}{\sqrt3},
\nonumber\\
\langle[N]\phi L;T,-T|\hat N_1|[N]\phi L;T,-T\rangle
&=&\frac{T}{\sqrt2},
\nonumber\\
\langle[N]\phi L;T,-T|\hat N_2|[N]\phi L;T,-T\rangle
&=&-\frac{T(2N+3)}{\sqrt{6}(2T+3)}.
\label{Aver}
\end{eqnarray}
Finally, the average boson numbers
are obtained through combination of the previous equations,
\begin{eqnarray}
\langle\hat N_\pi\rangle
&\equiv&
\langle[N]\phi L;T,-T|\hat N_\pi|[N]\phi L;T,-T\rangle
=\frac{(T+1)(N-T)}{2T+3},
\nonumber\\
\langle\hat N_\delta\rangle
&\equiv&
\langle[N]\phi L;T,-T|\hat N_\delta|[N]\phi L;T,-T\rangle
=\frac{N-T}{2T+3},
\nonumber\\
\langle\hat N_\nu\rangle
&\equiv&
\langle[N]\phi L;T,-T|\hat N_\nu|[N]\phi L;T,-T\rangle
=\frac{T(N+T)+(N+2T)}{2T+3}.
\label{Avera}
\end{eqnarray}
These expectation values satisfy the usual relations
\begin{equation} 
\langle\hat N_\pi\rangle+
\langle\hat N_\delta\rangle+
\langle\hat N_\nu\rangle=N,
\qquad
\langle\hat N_\nu\rangle-\langle\hat N_\pi\rangle=T.
\end{equation} 
The expressions (\ref{Avera})
do not depend on the orbital part of the state,
but only on its $SU_T(3)$ symmetry character.
Being valid for all symmetric states,
they are quite insensitive to the detailed structure
of a specific IBM-3 Hamiltonian.

\section{Predictions and Comparison with Experiment}
\label{sec-exp}
Within the context of IBM-1 and IBM-2
extensive calculations have been performed
for energy spectra and electromagnetic properties
of medium-mass and heavy nuclei,
generally yielding a satisfactory agreement with the data \cite{Iach87}.
The IBM-3 has been applied less extensively
and correspondingly less is known about its viability.
In addition, IBM-3 applications differ somewhat
from previous IBM-1 and IBM-2 studies,
and this in two respects.
Firstly, its region of applicability
is essentially confined to $N\approx Z$ nuclei
(the only nuclei where proton--neutron pairs
might play a role at low energies)
and hence to lighter nuclei
which generally exhibit less collectivity
than those with higher mass number.
The IBM-3 can only provide a partial description of such nuclei
and inevitably misses out a number of levels
(of single-particle or intruder character) at low energies.
Secondly, because it conserves the isospin quantum number,
the IBM-3 can be linked (more naturally than either IBM-1 or IBM-2)
to the shell-model
and previous IBM-3 studies have been concerned primarily
with the connection between IBM-3 and the shell model,
rather than with phenomenological applications.
In particular, IBM-3 calculations
have been performed in the $f_{7/2}$ and $pf$ shells
\cite{Abde88,Abde89,Elli96,Thom87}
with Hamiltonians and electromagnetic operators
derived from a shell-model mapping.
The agreement with experimental observations
is good enough for low-lying symmetric states,
but for higher-lying states discrepancies occur.
An additional problem is
that the number of data points for non-symmetric states is quite low.

As far as phenomenological applications
of IBM-3 to $N\approx Z\approx40$ nuclei are concerned,
an example can be found in Ref.~\cite{Sugi97}
where a schematic Hamiltonian is taken
with parameters obtained
through a fit to energy spectra of several nuclei.
As an illustration of this type of approach
we use here a schematic Hamiltonian
with a ($d$-boson) pairing term
and a quadrupole--quadrupole interaction,
and apply it to the $f_{7/2}$ nuclei
$^{44}$Ti, $^{46}$Ti, $^{48}$Ti, and $^{48}$Cr.
For simplicity no dependence on $N$ and $T$ is considered
in the Hamiltonian parameters
and the quadrupole force is assumed to act exclusively
between protons and neutrons,
and not between identical particles \cite{Sugi97}. 
The Hamiltonian then reads
\begin{equation}
\hat H=
\epsilon_d\hat n_d+
\kappa_0 {\cal N}[\hat T^{(2,0)}:\hat T^{(2,0)}+
{2\over3}\hat T^{(2,1)}:\hat T^{(2,1)}]+
t\hat T^2,
\label{sch-ham}
\end{equation}
where the symbol $:$
denotes a scalar product in isospin and orbital space
and $\cal N[\dots]$ stands for a normal-ordered product.
The structure parameters of the quadrupole operators
are assumed to be independent of $T$:
$\chi^{2T}_{02}=\chi^{2T}_{20}=1$
and $\chi^{2T}_{22}=\chi$.
The parameter $\epsilon_d$ is obtained
from the binding energies and single-particle levels of
$^{40-42}$Ca, $^{41-42}$Sc, and $^{42}$Ti.
The parameters $\kappa_0$ and $\chi$
are determined through a best fit to low-energy states
in $^{44,46,48}$Ti and $^{48}$Cr.
Finally, $t$ is obtained
through a comparison with a shell-model calculation \cite{Thom87}
since reliable experimental information on states with $T>|M_T|$
is scarce.
The parameters thus determined are:
$\epsilon_d=1.5$ MeV (the same for $\pi$, $\delta$, and $\nu$ bosons),
$\kappa_0=-0.2$ MeV,
$\chi=-2.4$,
and $t=1.2$ MeV.
It is worth noting
that the quadrupole--quadrupole interaction
contains a sizeable contribution to the Majorana term \cite{Sugi97}
and this guarantees an approximate $U_L(6)$ symmetry
for the low-lying states.
The experimental and theoretical spectra are shown in Fig.~\ref{comp-spec}.
An overall agreement is observed for the low-lying states of all nuclei;
the degree of agreement is similar to the one obtained in Ref.~\cite{Elli96}.

Regarding the electromagnetic transitions,
some general conclusions
independent of the $U_L(6)$ symmetry can be obtained. 
Unfortunately there are only few measured $B$(M1) transition rates
in this mass region
and it is difficult to determine
the parameters in the electromagnetic operators.
In Table~\ref{tab-comp-ener} the observed energies of non-symmetric states
are compared to the ones calculated with IBM-3,
the non-symmetric character of the states
being proposed on the basis of their decay properties.
In Table~\ref{tab-comp-BM1} the $B$(M1) values in IBM-3
are compared to the observed ones when known
or to the shell-model values otherwise.
Two parameters enter the IBM-3 calculation:
the isovector boson $g$ factor
which is fixed to reproduce the $B({\rm M1};2^+_M\rightarrow2^+_1)$ value
in $^{44}$Ti $(g_1=1.20\mu_N)$.
Although not suggested by microscopy,
one may study the influence of an isotensor boson $g$ factor
which is also illustrated in Table~\ref{tab-comp-BM1} with $(g_2=0.58\mu_N)$,
derived from the $B({\rm M1};2^+_M\rightarrow2^+_1)$ in $^{46}$Ti.
Rather unexpectedly,
it is seen that the isotensor contribution improves the fit significantly.
Note that the isotensor contribution for $^{44}$Ti and $^{48}$Cr vanishes
because $T_z=0$ for both nuclei.

Although boson-number expectation values are
not directly measurable quantities,
a comparison with more elaborate calculations can be carried out.
In Fig.~\ref{bos-num}
the expectation value of boson numbers in IBM-3
in the ground state of even--even Fe and Cr isotopes
is compared with SMMC calculations \cite{Enge96}.
It is seen that the simple formulas (\ref{Avera})
qualitatively reproduce the features of full microscopic calculations.
The SMMC results are scaled
such that the pair-number expectation values
are normalized to the total number of bosons,
which is only approximately valid for shells with finite size.

The contributions of the Majorana and $\hat T^2$ terms in the Hamiltonian
are crucial to the relative position of states
with different $U_L(6)$ symmetry and isospin.
We illustrate this with a schematic calculation
that is relevant for $N=Z$ nuclei.
It is generally assumed
that states with $[N-2,2]$ $U_L(6)$ symmetry
occur at higher energy than those with $[N-1,1]$.
This, however, is not necessarily so in $N=Z$ even--even nuclei.
This peculiarity arises
because the lowest allowed isospin value
in the $[N-1,1]$ representation is $T=1$,
while it can be $T=0$ in the $[N-2,2]$ representation.
Thus, although the Majorana term
always favors $[N-1,1]$ over $[N-2,2]$ states,
in $N=Z$ nuclei this effect is counteracted by the $\hat T^2$ term.
To make the argument somewhat more quantitative,
assume a simple Hamiltonian
with a $\hat T^2$ and a Majorana term
besides a general orbital dependence $\hat H_L$,
\begin{equation}
\hat H=t\hat T^2+m\hat M+\hat H_L.
\label{sch-ham2}
\end{equation}
Appropriate values of the parameters in the $f_{7/2}$ shell
are $t=1.2$ MeV and $m=3.3$, 2.5, and 0.7 MeV
for $N=2$, 3, and 4, respectively,
where the values of $m$ are estimated
from the energies of non-symmetric states in $^{44,46,48}$Ti
\cite{Abde88,Abde89}.
Application of the schematic Hamiltonian (\ref{sch-ham2})
to an $N=Z$ nucleus with four bosons
(e.g., $^{48}$Cr or $^{64}$Ge)
gives the result shown in Fig.~\ref{sch-spec}.
It is seen in particular
that the $[N-2,2]T=0$ states
occur at a lower energy than those with $[N-1,1]T=1$.
An intruiging feature is
that the $[N-2,2]$ states cannot decay
to the symmetric ground-state configuration
if one assumes the electromagnetic operators
of one-body type in the bosons.
However, this effect is not very likely to persist
for realistic Hamiltonians because of $U_L(6)$ mixing
between $[N-2,2]$ and $[N-1,1]$ states.

\section{Conclusions}
\label{sec-conclu}
The main objective of this paper
was to present a comprehensive analysis
of the symmetry limits of the IBM-3
that conserve $SU_T(3)$ charge or $U_L(6)$ $sd$ symmetry.
Although particular results were obtained
for the three limits $U_L(5)$, $SU_L(3)$, and $O_L(6)$,
special emphasis was given to a general analysis
independent of the latter limits
but only requiring $SU_T(3)$ or $U_L(6)$ symmetry.
The origin of this symmetry was shown
to be related to the Majorana interaction in the IBM-3 Hamiltonian
which leads to a decoupling of the orbital and isospin spaces.
In previous, microscopic studies of IBM-3
the $SU_T(3)$ charge symmetry had been shown to be approximately valid
and this paper has taken this result as a starting point
to derive the properties of all limits with that symmetry.

A numerical application of the IBM-3 was presented
involving a simple, phenomenological Hamiltonian
with a few parameters either derived from shell-model considerations
or fitted to the data.
This Hamiltonian was applied to even--even $f_{7/2}$ nuclei
where the $SU_T(3)$ charge symmetry is thought to be approximately valid.
Reasonable results were obtained
but any analysis of this kind
will ultimately be hampered by the limited amount of collectivity
exhibited by these nuclei.
Applications to regions of more collective nuclei
should thus be considered (e.g., $28\leq N,Z\leq50$).

A valuable aspect of algebraic models
is that they usually are simple enough
as to give clues concerning key observable quantities.
The IBM-3 is no exception.
Examples are the expressions derived
for the expectation values of the various boson numbers;
although these cannot be measured,
they can be compared to similar expectation values
calculated in more elaborate models.
In particular, the IBM-3 expectation values were shown
to be in good agreement with Shell Model Monte Carlo calculations.
The present analysis has also revealed
two intriguing predictions of the IBM-3.
The first concerns the transfer probility of a $\delta$ boson
(i.e., a proton--neutron $T=1$ pair)
which turns out to be zero in IBM-3.
As such, this result is too schematic to assign it too much weight:
the transfer necessarily takes place from or into an odd--odd nucleus
for which an IBM-3 description is incomplete.
However, it indicates that the same problem should be revisited in IBM-4
where it possibly can teach us something
about the proton--neutron pair structure of $N\sim Z$ nuclei.
The second intriguing prediction of the IBM-3
concerns the energy of non-symmetric states.
In $N=Z$ nuclei, and in $N=Z$ nuclei {\em only},
it is conceivable that the $[N-2,2]$ states occur at a lower energy
than those with $[N-1,1]$ symmetry.
Assuming exact $SU_T(3)$ charge symmetry,
the decay from $[N-2,2]$ to the symmetric ground-state configuration $[N]$
is forbidden.
The breaking of $SU_T(3)$ charge symmetry
will, however, destroy this selection rule;
it would nevertheless be of interest to verify
whether any remnant of it is still observable.

\section*{Acknowledgment}
This work has been supported in part by the Spanish DGICYT under contract
No. PB95--0533, by the European Commission under contract CI1*-CT94-0072
and by a IN2P3 (France)-CICYT (Spain) agreement.

\begin{table}
\caption{Lowest Eigenstates of a $U(5)$ Hamiltonian}
\label{tab-sta-u5}
\begin{displaymath}
\begin{array}{lcl}
\hline\hline\\
| 0_1^+;T\rangle&=&|[N](0,0,0)(0,0)0;T\rangle 
\\
| 0_2^+(d^2);T\rangle&=&|[N](2,0,0)(0,0)0;T\rangle 
\\
| 0_3^+(d^3);T\rangle&=&|[N](3,0,0)(3,0)0;T\rangle 
\\
| 1_M^+(d^2);T\rangle&=&|[N-1,1](1,1,0)(1,1)1;T\rangle 
\\
| 2_1^+(d);T\rangle&=&|[N](1,0,0)(1,0)2;T\rangle
\\
| 2_2^+(d^2);T\rangle&=&|[N](2,0,0)(2,0)2;T\rangle
\\
| 2_3^+(d^3);T\rangle&=&|[N](3,0,0)(1,0)2;T\rangle
\\
| 2_M^+(d);T\rangle&=&|[N-1,1](1,0,0)(1,0)2;T\rangle
\\
| 3_1^+(d^3);T\rangle&=&|[N](3,0,0)(3,0)3;T\rangle
\\
| 3_M^+(d^2);T\rangle&=&|[N-1,1](1,1,0)(1,1)3;T\rangle
\\
| 4_1^+(d^2);T\rangle&=&|[N](2,0,0)(2,0)4;T\rangle
\\
| 4_2^+(d^3);T\rangle&=&|[N](3,0,0)(3,0)4;T\rangle
\\ \\
\hline\hline
\end{array}
\end{displaymath}
\end{table}

\begin{table}
\caption{Lowest Eigenstates of an $SU(3)$ Hamiltonian}
\label{tab-sta-su3}
\begin{displaymath}
\begin{array}{lcl}
\hline\hline\\
| 0_1^+;T\rangle&=&|[N](2N,0)0;T\rangle
\\
| 0_{\beta}^+;T\rangle&=&|[N](2N-4,2)2;T\rangle
\\
| 1_M^+;T\rangle&=&|[N-1,1](2N-2,1)1;T\rangle
\\
| 2_1^+;T\rangle&=&|[N](2N,0)2;T\rangle
\\
| 2_{\beta}^+;T\rangle&=&|[N](2N-4,2)02;T\rangle
\\
| 2_{\gamma}^+;T\rangle&=&|[N](2N-4,2)22;T\rangle
\\
| 2_M^+;T\rangle&=&|[N-1,1](2N-2,1)2;T\rangle
\\
| 3_{\gamma}^+;T\rangle&=&|[N](2N-4,2)3;T\rangle
\\
| 3_M^+;T\rangle&=&|[N-1,1](2N-2,1)3;T\rangle
\\
| 3_{M'}^+;T\rangle&=&|[N-1,1](2N-4,2)3;T\rangle
\\
| 4_1^+;T\rangle&=&|[N](2N,0)4;T\rangle
\\
| 4_{\beta}^+;T\rangle&=&|[N](2N-4,2)04;T\rangle
\\
| 4_{\gamma}^+;T\rangle&=&|[N](2N-4,2)24;T\rangle
\\ \\
\hline\hline
\end{array}
\end{displaymath}
\end{table}

\begin{table}
\caption{Lowest Eigenstates of an $O(6)$ Hamiltonian}
\label{tab-sta-o6}
\begin{displaymath}
\begin{array}{lcl}
\hline\hline\\
| 0_1^+;T\rangle&=&|[N]( N,0,0 \rangle
(0,0)0;T \rangle 
\\ 
| 0_2^+;T\rangle&=&|[N](N,0,0)(3,0)0;T\rangle 
\\
| 0_3^+;T\rangle&=&|[N](N-2,0,0)(0,0)0;T\rangle 
\\
| 1_M^+;T\rangle&=&|[N-1,1](N-1,1,0)(1,1)1;T\rangle 
\\
| 2_1^+;T\rangle&=&|[N](N,0,0)(1,0)2;T\rangle 
\\
| 2_2^+;T\rangle&=&|[N](N,0,0)(2,0)2;T\rangle 
\\
| 2_3^+;T\rangle&=&|[N](N-2,0,0)(1,0)2;T\rangle 
\\
| 2_4^+;T\rangle&=&|[N](N-2,0,0)(2,0)2;T\rangle 
\\
| 2_M^+;T\rangle&=&|[N-1,1](N-1,1,0)(1,0)2;T\rangle 
\\
| 3_1^+;T\rangle&=&|[N](N,0,0)(3,0)3;T\rangle 
\\
| 3_M^+;T\rangle&=&|[N-1,1](N-1,1,0)(1,1)3;T\rangle 
\\
| 4_1^+;T\rangle&=&|[N](N,0,0)(2,0)4;T\rangle 
\\
| 4_2^+;T\rangle&=&|[N](N-2,0,0)(2,0)4;T\rangle 
\\ \\
\hline\hline
\end{array}
\end{displaymath}
\end{table}

\begin{table}
\caption{E2 Excitation out of the Ground State for the $U(5)$ Limit}
\label{tab-tran-u5}
\begin{displaymath}
\begin{array}{cccccc}
\hline\hline
J^{\pi}_{\rm f} && \cal{T}\lambda && T 
& B({\cal{T}} \lambda,T;0_1^+\rightarrow J^{\pi}_{\rm f})
\\ 
\hline\\ 
2_1^+     && E2     && 0 & (e_{\pi}+e_{\delta}+e_{\nu})^2 \,5N     \\     \\ 
2_1^+     && E2     && 1 & (e_{\pi}-e_{\nu})^2\, 
                           {\displaystyle\frac{5T^2}{N}}       \\ \\ 
2_1^+     && E2     && 2 & (-e_{\pi}+2e_{\delta}-e_{\nu})^2\,
                           {\displaystyle\frac{5T^2(2N+3)^2}{N(2T+3)^2}} \\ \\
2_M^+     && E2     && 1 & (e_{\pi}-e_{\nu})^2\;
                           {\displaystyle\frac{5T(N-T)(N+T+1)}{N(T+1)}}   \\ \\
2_M^+     && E2     && 2 & (-e_{\pi}+2e_{\delta}-e_{\nu})^2\:
                           {\displaystyle\frac{45T(N-T)
                           (N+T+1)}{N(T+1)(2T+3)^2}}           \\    \\
\hline\hline
\end{array}
\end{displaymath}
\end{table}

\begin{table}
\caption{M1 and E2 Excitation out of the Ground State for the $SU(3)$ Limit}
\label{tab-tran-su3-1}
\begin{displaymath}
\begin{array}{cccccc}
\hline\hline
J^{\pi}_{\rm f} && \cal{T}\lambda && T
& B({\cal{T}}\lambda,T;0_1^+\rightarrow J^{\pi}_{\rm f})
\\ 
\hline \\
1_M^+     && M1     && 1 & {\displaystyle
                           \frac{3}{4\pi}(g_{\pi}-g_{\nu})^2 \frac{8T(N-T)
                           (N+T+1)}{(2N-1)(T+1)}}                         \\ \\
1_M^+     && M1     && 2 & {\displaystyle
                           \frac{3}{4\pi}(-g_{\pi}+2g_{\delta}-g_{\nu})^2
                           \frac{72T(N-T)(N+T+1)}{(T+1)(2T+3)^2(2N-1)}}   \\ \\
2_1^+     && E2     && 0 & {\displaystyle
                           (e_{\pi}+e_{\delta}+e_{\nu})^2 N(2N+3)}        \\ \\
2_1^+     && E2     && 1 & {\displaystyle
                           (e_{\pi}-e_{\nu})^2 \frac{T^2(2N+3)}{N}}       \\ \\
2_1^+     && E2     && 2 & {\displaystyle
                           (-e_{\pi}+2e_{\delta}-e_{\nu})^2\frac{T^2(2N+3)^3}
                           {N(2T+3)^2}}                                   \\ \\
2_M^+     && E2     && 1 & {\displaystyle
                           (e_{\pi}-e_{\nu})^2 \frac{3T(N-1)(N-T)(N+T+1)}
                           {N(2N-1)(T+1)}}                                \\ \\
2_M^+     && E2     && 2 & {\displaystyle
                           (-e_{\pi}+2e_{\delta}-e_{\nu})^2\frac{27T(N-T)(N-1)
                           (N+T+1)}{N(2N-1)(T+1)(2T+3)^2}}                \\ \\
\hline\hline
\end{array}
\end{displaymath}
\end{table}

\begin{table}
\caption{M3 Excitation out of the Ground State for the $SU(3)$ Limit}
\label{tab-tran-su3-2}
\begin{displaymath}
\begin{array}{cccccc}
\hline\hline
J^{\pi}_{\rm f} && \cal{T}\lambda && T
& B({\cal{T}}\lambda,T;0_1^+\rightarrow J^{\pi}_{\rm f})
\\
\hline \\
3_{\gamma}^+&&M3    && 0 & {\displaystyle
                           \frac{35}{8\pi}(\Omega_{\pi}+\Omega_{\delta}+
                           \Omega_{\nu})^2\frac{8N(N-2)(N-1)}{3(2N-3)(2N-1)}}
                                                                          \\ \\
3_{\gamma}^+&&M3    && 1 & {\displaystyle
                           \frac{35}{8\pi}(\Omega_{\pi}-\Omega_{\nu})^2 
                           \frac{8(N-2)(N-1)T^2}{3N(2N-1)(2N-3)}}         \\ \\
3_{\gamma}^+&&M3    && 2 & {\displaystyle
                           \frac{35}{8\pi}(-\Omega_{\pi}+2\Omega_{\delta}-
                           \Omega_{\nu})^2\frac{8T^2(N-1)(N-2)(2N+3)^2}
                           {3N(2N-1)(2N-3)(2T+3)^2}}                      \\ \\
3_M^+     && M3     && 1 & {\displaystyle
                           \frac{35}{8\pi}(\Omega_{\pi}-\Omega_{\nu})^2 
                           \frac{4T(N-T)(2N+3)(N+T+1)}{15(N-1)(2N-1)(T+1)}}
                                                                         \\ \\
3_M^+     && M3     && 2 & {\displaystyle
                           \frac{35}{8\pi}(-\Omega_{\pi}+2\Omega_{\delta}-
                           \Omega_{\nu})^2\frac{12T(N-T)(2N+3)(N+T+1)}
                           {(T+1)(2T+3)^2(N-1)(2N-1)}}                    \\ \\
3_{M'}^+  && M3     && 1 & {\displaystyle
                           \frac{35}{8\pi}(\Omega_{\pi}-\Omega_{\nu})^2 
                           \frac{4T(N-T)(N-2)^2(N+T+1)}{3N(N-1)(2N-3)(T+1)}}  
                                                                          \\ \\
3_{M'}^+  && M3     && 2 & {\displaystyle
                           \frac{35}{8\pi}(-\Omega_{\pi}+2\Omega_{\delta}-
                           \Omega_{\nu})^2\frac{12T(N-T)(N-2)^2(N+T+1)}
                           {N(N-1)(2N-3)(T+1)(2T+3)^2}}                   \\ \\
\hline\hline
\end{array}
\end{displaymath}
\end{table}

\begin{table}
\caption{M1, E2, and M3 Excitation out of the Ground State
for the $O(6)$ Limit}
\label{tab-tran-o6}
\begin{displaymath}
\begin{array}{cccccc}
\hline\hline
J^{\pi}_{\rm f} && \cal{T}\lambda && T
& B({\cal{T}}\lambda,T;0_1^+\rightarrow J^{\pi}_{\rm f})
\\ 
\hline \\
1_M^+     && M1     && 1 & {\displaystyle
                           \frac{3}{4\pi}(g_{\pi}-g_{\nu})^2 \frac{3T(N-T)
                           (N+T+1)}{(N+1)(T+1)}}                          \\ \\
1_M^+     && M1     && 2 & {\displaystyle
                           \frac{3}{4\pi}(-g_{\pi}+2g_{\delta}-g_{\nu})^2
                           \frac{27T(N-T)(N+T+1)}{(T+1)(2T+3)^2(N+1)}}    \\ \\
2_1^+     && E2     && 0 & {\displaystyle
                           (e_{\pi}+e_{\delta}+e_{\nu})^2 N(N+4)}         \\ \\
2_1^+     && E2     && 1 & {\displaystyle
                           (e_{\pi}-e_{\nu})^2 \frac{T^2(N+4)}{N}}        \\ \\
2_1^+     && E2     && 2 & {\displaystyle
                           (-e_{\pi}+2e_{\delta}-e_{\nu})^2\frac{T^2(N+4)
                           (2N+3)^2}{N(2T+3)^2}}                          \\ \\
2_M^+     && E2     && 1 & {\displaystyle
                           (e_{\pi}-e_{\nu})^2 \frac{2T(N+2)(N-T)(N+T+1)}
                           {N(N+1)(T+1)}}                                 \\ \\
2_M^+     && E2     && 2 & {\displaystyle
                           (-e_{\pi}+2e_{\delta}-e_{\nu})^2\frac{18T(N-T)(N+2)
                           (N+T+1)}{N(N+1)(T+1)(2T+3)^2}}                 \\ \\
3_M^+     && M3     && 1 & {\displaystyle
                           \frac{35}{8\pi}(\Omega_{\pi}-\Omega_{\nu})^2 
                           \frac{7T(N-T)(N+T+1)}{10(N+1)(T+1)}}           \\ \\
3_M^+     && M3     && 2 & {\displaystyle
                           \frac{35}{8\pi}(-\Omega_{\pi}+2\Omega_{\delta}-
                           \Omega_{\nu})^2\frac{63T(N-T)(N+T+1)}
                           {10(T+1)(2T+3)^2(N+1)}}                        \\ \\
\hline\hline
\end{array}
\end{displaymath}
\end{table}

\begin{table}
\caption{Two-Neutron Transfer Intensities in the $SU(3)$ Limit}
\label{tab-two-nuc}
\begin{displaymath}
\begin{array}{ll}
\hline\hline\\
I([N]0^+_1;T,-T \rightarrow [N+1]0^+_1;T-1,-T+1)&=p_{\pi,0}^2
{\displaystyle
\frac{T(N-T+2)(2N+3)}{3(2T+1)(2N+1)}}
\\ \\
I([N](L-2)^+_1;T,-T \rightarrow [N+1]L^+_1;T-1,-T+1)&=p_{\pi,2}^2
{\displaystyle
\frac{L(L-1)(2N+L+1)}{(2L-3)(2L-1)(2T+1)}}
\\ \\
& \times {\displaystyle
\frac{(2N+L+3)T(N-T+2)}{(2N+1)(2N+2)}}
\\ \\
I([N]0^+_1;T,-T \rightarrow [N+1] 0^+_1;T,-T)&=0
\\ \\
I([N](L-2)^+_1;T,-T \rightarrow [N+1]L^+_1;T,-T)&=0
\\ \\
I([N] 0^+_1;T,-T \rightarrow [N+1] 0^+_1;T+1,-T-1)&=p_{\nu,0}^2
{\displaystyle
\frac{(T+1)(2N+3)(N+T+3)}{3(2T+3)(2N+1)}}
\\ \\
I([N](L-2)^+_1;T,-T \rightarrow [N+1]L^+_1;T+1,-T-1)&=p_{\nu,2}^2
{\displaystyle
\frac{L(L+1)(2N+L+1)}
{(2L-3)(2L-1)}}
\\ \\
& \times {\displaystyle
\frac{(2N+L+3)(T+1)(N+T+3)}{(2T+3)(2N+1)(2N+2)}}
\\ \\
\hline\hline
\end{array}
\end{displaymath}
\end{table}

\begin{table}
\caption{Energies of Non-Symmetric States in $^{44,46,48}$Ti and $^{48}$Cr}
\label{tab-comp-ener}
\begin{tabular}{ccll}
Nucleus & State & \multicolumn{2}{c}{Energy (MeV)}\\
\cline{3-4}
        &       & Observed & IBM-3\\
\hline
$^{44}$Ti & 1$_1^+$ & 5.7$^a$ & 5.2 \\
          & 2$_M^+$ & 6.6     & 4.8 \\
$^{46}$Ti & 1$_1^+$ & 4.3     & 2.8 \\
          & 2$_M^+$ & 2.5$^a$ & 2.1 \\
          & 3$_M^+$ & 3.6$^a$ & 3.8 \\
$^{48}$Ti & 1$_1^+$ & 3.7     & 2.9 \\
          & 2$_M^+$ & 2.4     & 2.2 \\
          & 3$_M^+$ & 3.2     & 4.3 \\
$^{48}$Cr & 1$_1^+$ & 5.5$^a$ & 5.4 \\
\end{tabular}
$^a$ Calculated value from Refs.~\cite{Abde88,Abde89}.
\end{table}

\newpage
\begin{table}
\caption{M1 Transition Rates between Symmetric and Non-Symmetric States
in $^{44,46,48}$Ti and $^{48}$Cr}
\label{tab-comp-BM1}
\begin{tabular}{cclll}
Nucleus & Transition & \multicolumn{3}{c}{$B$(M1) ($\mu_{\rm N}^2$)} \\
\cline{3-5}
       &            & Observed & Shell model$^a$ & IBM-3$^b$\\
\hline
$^{44}$Ti & 2$_M^+\rightarrow$2$_1^+$ & & 1.14 &1.14 (1.14) \\
          & 0$_1^+\rightarrow$1$_1^+$ & & 2.40 &1.75 (1.75) \\
$^{46}$Ti & 2$_M^+\rightarrow$2$_1^+$ & & 0.73 &0.73 (1.13) \\ 
          & 3$_M^+\rightarrow$2$_1^+$ & & 0.07 &0.20 (0.25) \\ 
          & 3$_M^+\rightarrow$4$_1^+$ & & 0.20 &0.41 (0.69) \\ 
          & 0$_1^+\rightarrow$1$_1^+$ & 1.01&  &1.15 (1.34) \\ 
$^{48}$Ti & 2$_M^+\rightarrow$2$_1^+$ &0.50(10) & 0.58  & 0.90 (1.24) \\ 
          & 3$_M^+\rightarrow$2$_1^+$ &0.08(3)  & 0.003 & 0.30 (0.34) \\ 
          & 3$_M^+\rightarrow$4$_1^+$ &0.42(16) & 0.32  &0.49  (0.64) \\ 
          & 4$_M^+\rightarrow$4$_1^+$ &1.4(5)   & 1.50  &             \\ 
          & 0$_1^+\rightarrow$1$_1^+$ &0.50(8)  & 0.54  &1.82  (2.10) \\ 
$^{48}$Cr & 0$_1^+\rightarrow$1$_1^+$ & & 3.05 &4.82 (4.82) \\
\end{tabular}
\end{table}
$^a$ From Refs.~\cite{Abde88,Abde89,Caur94}. 

$^b$ This work.
The numbers in parentheses are without isotensor contribution.

\begin{figure}[hbt]
\begin{center}
\end{center}
\caption{Energy spectrum of $T=0$ and $T=1$ states
for a $U(5)$ Hamiltonian (\ref{HamU5}) with parameters (in MeV)
$A_2=-0.175$, $B_1=0.400$, $F_2=0.010$, and $\beta_2=1.2$.
The boson number is $N=4$.}
\label{u5-spec}
\end{figure}

\begin{figure}[]
\begin{center}
\end{center}
\caption{Energy spectrum of $T=0$ and $T=1$ states
for an $SU(3)$ Hamiltonian (\ref{HamSU3}) with parameters (in MeV)
$A_2=-0.175$, $C_2=-0.006$, $F_2=0.010$, and $\beta_2=1.2$.
The boson number is $N=4$.}
\label{su3-spec}
\end{figure}

\begin{figure}[]
\begin{center}
\end{center}
\caption{Energy spectrum of $T=0$ and $T=1$ states
for an $O(6)$ Hamiltonian (\ref{HamO6}) with parameters (in MeV)
$A_2=-0.175$, $D_2=-0.035$, $E_2=0.035$, $F_2=0.010$, and $\beta_2=1.2$.
The boson number is $N=4$.}
\label{o6-spec}
\end{figure}

\begin{figure}[]
\begin{center}
\end{center}
\caption{Comparison of experimental and IBM-3 excitation energies
of $^{44,46,48}$Ti and $^{48}$Cr.
The parameters are constant for all isotopes
and are given in the text.}
\label{comp-spec}
\end{figure}

\begin{figure}[]
\begin{center}
\end{center}
\caption{Ground-state boson-number expectation values
in Cr and Fe isotopes
as a function of $|T_z|=(N-Z)/2$
calculated with SMMC (left) and IBM-3 (right).}
\label{bos-num}
\end{figure}

\begin{figure}[]
\begin{center}
\end{center}
\caption{Energies of [4], [3,1], and [2,2] states
in an $N=Z$ nucleus with four bosons.
The schematic Hamiltonian (\ref{sch-ham2}) is used with parameters (in MeV)
$t=1.2$ and $m=0.7$.
Transitions between [4] and [2,2] (dashed line) are forbidden.}
\label{sch-spec}
\end{figure}

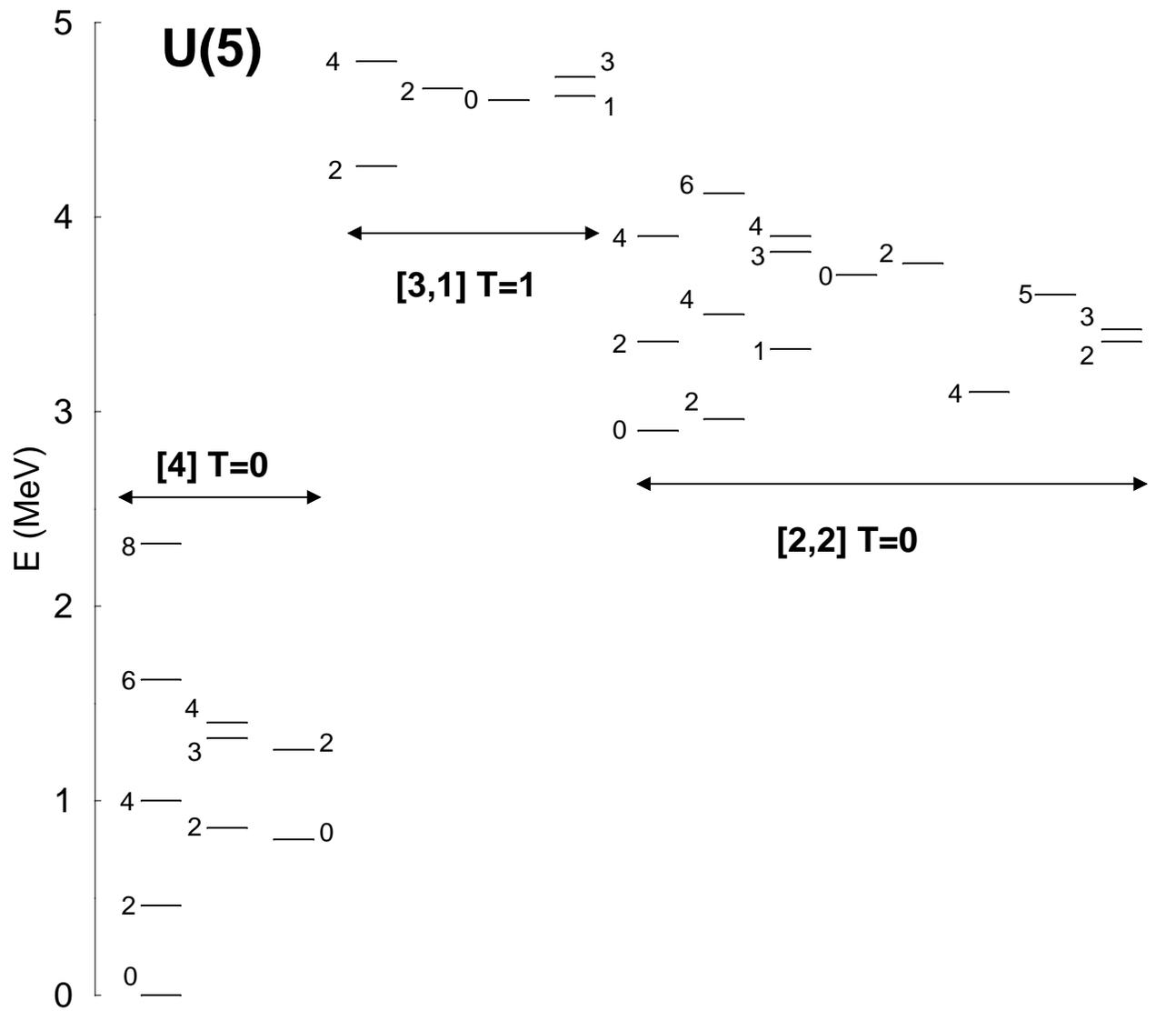
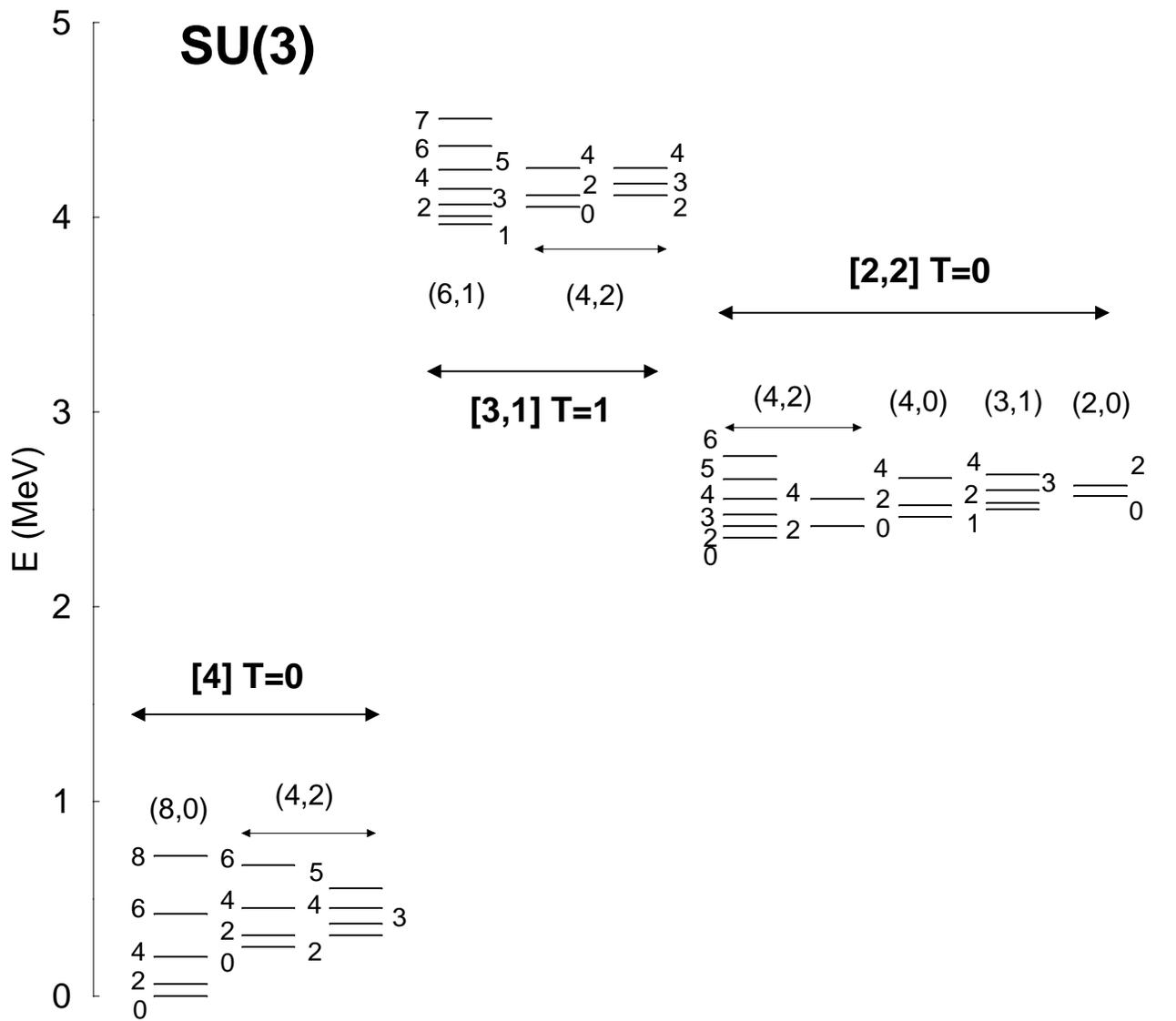
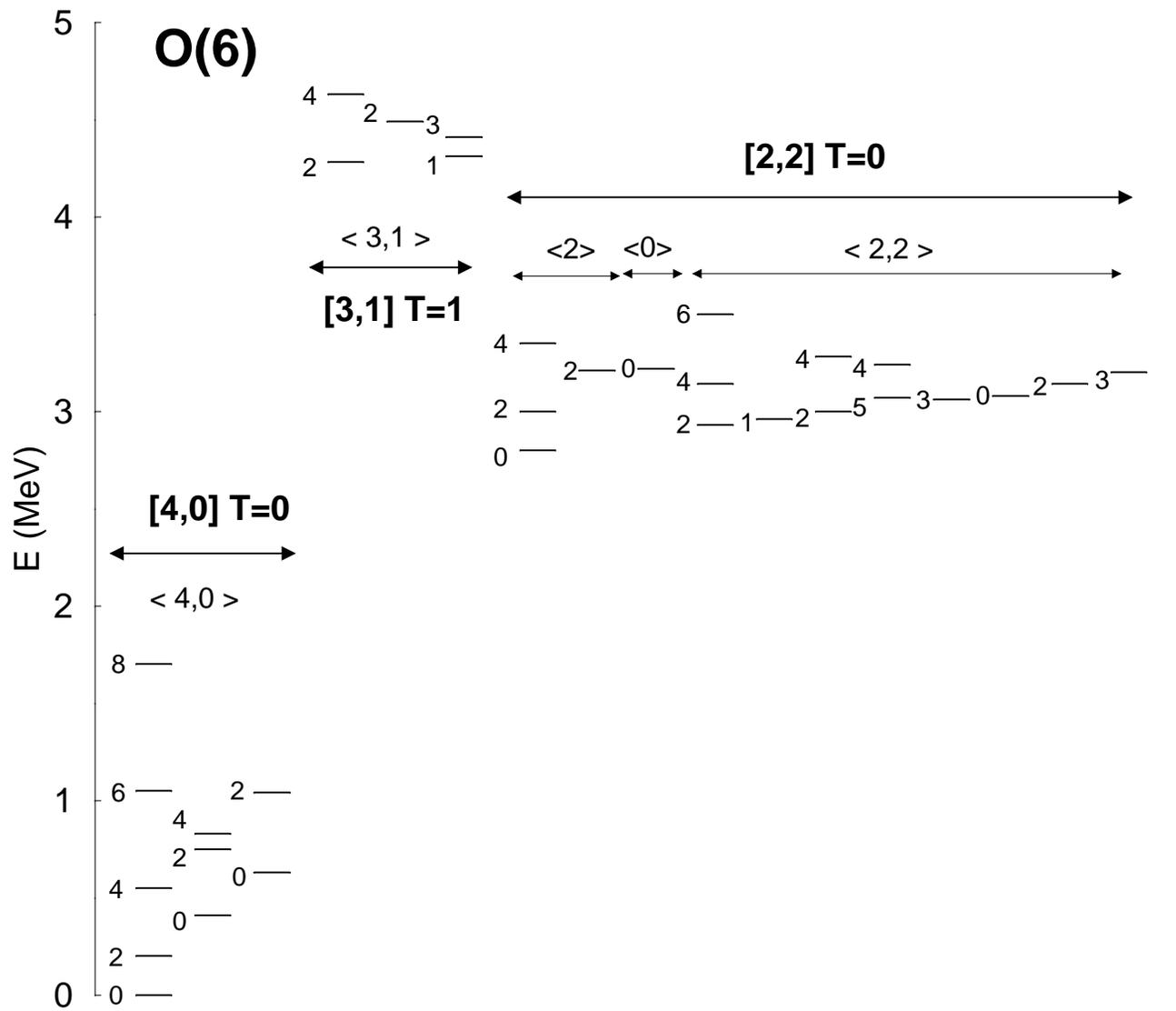
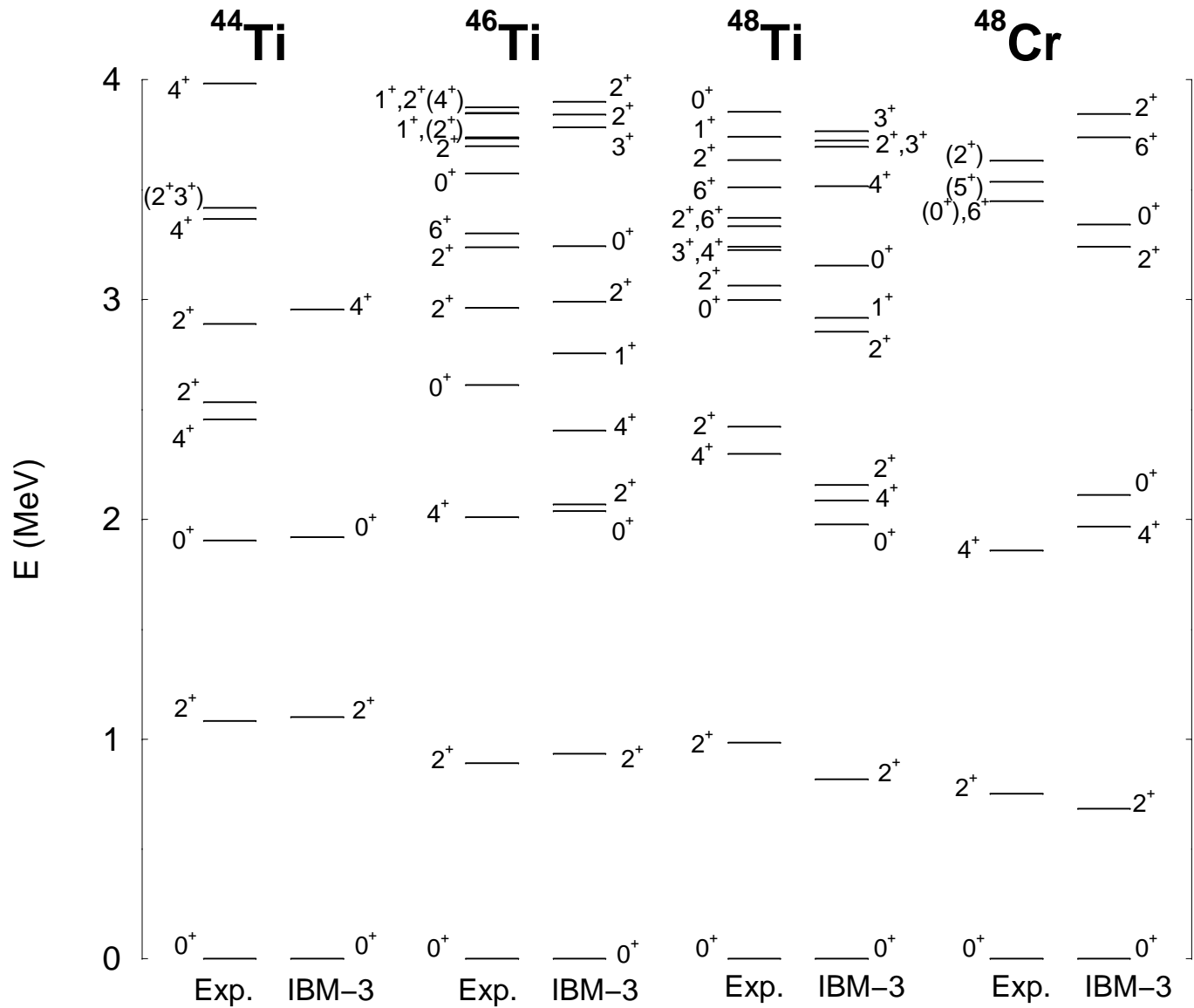
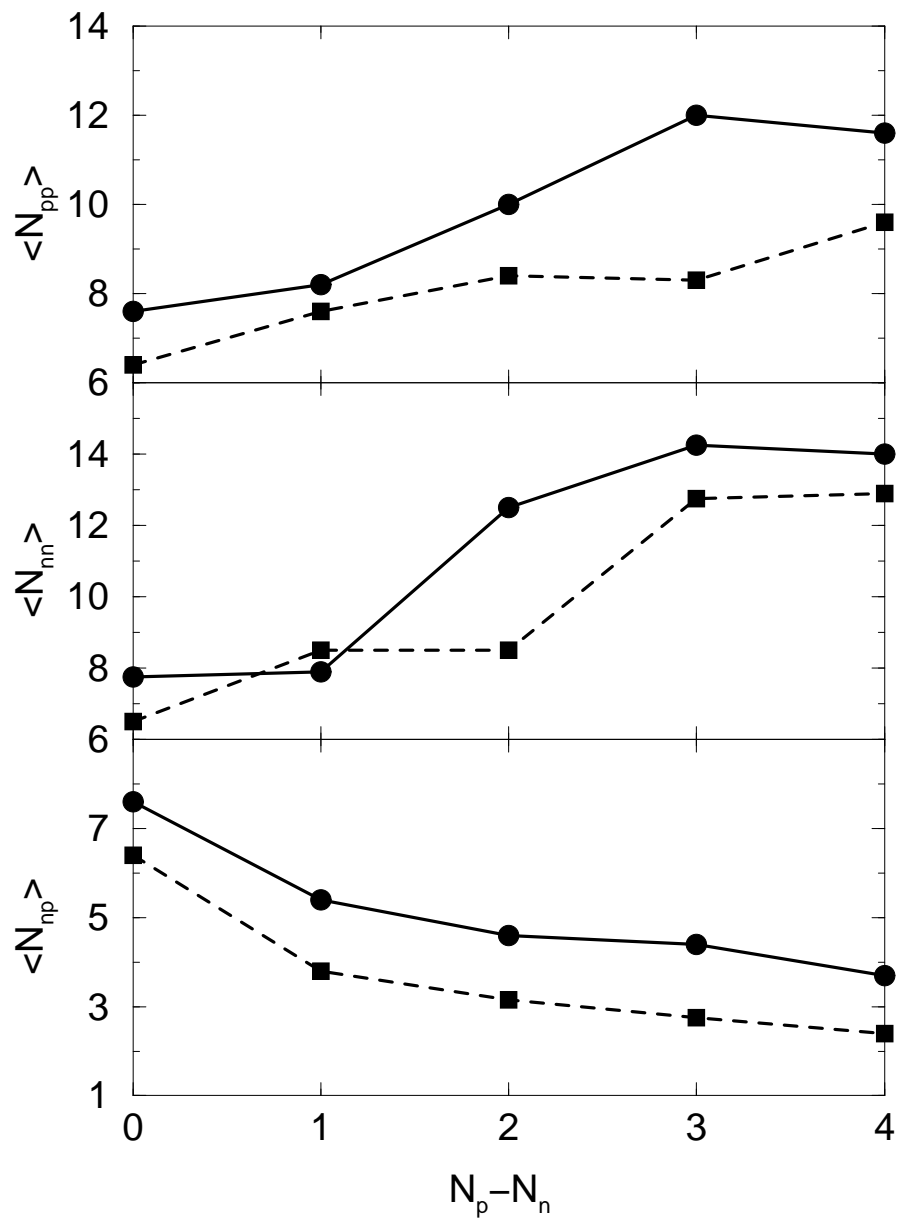
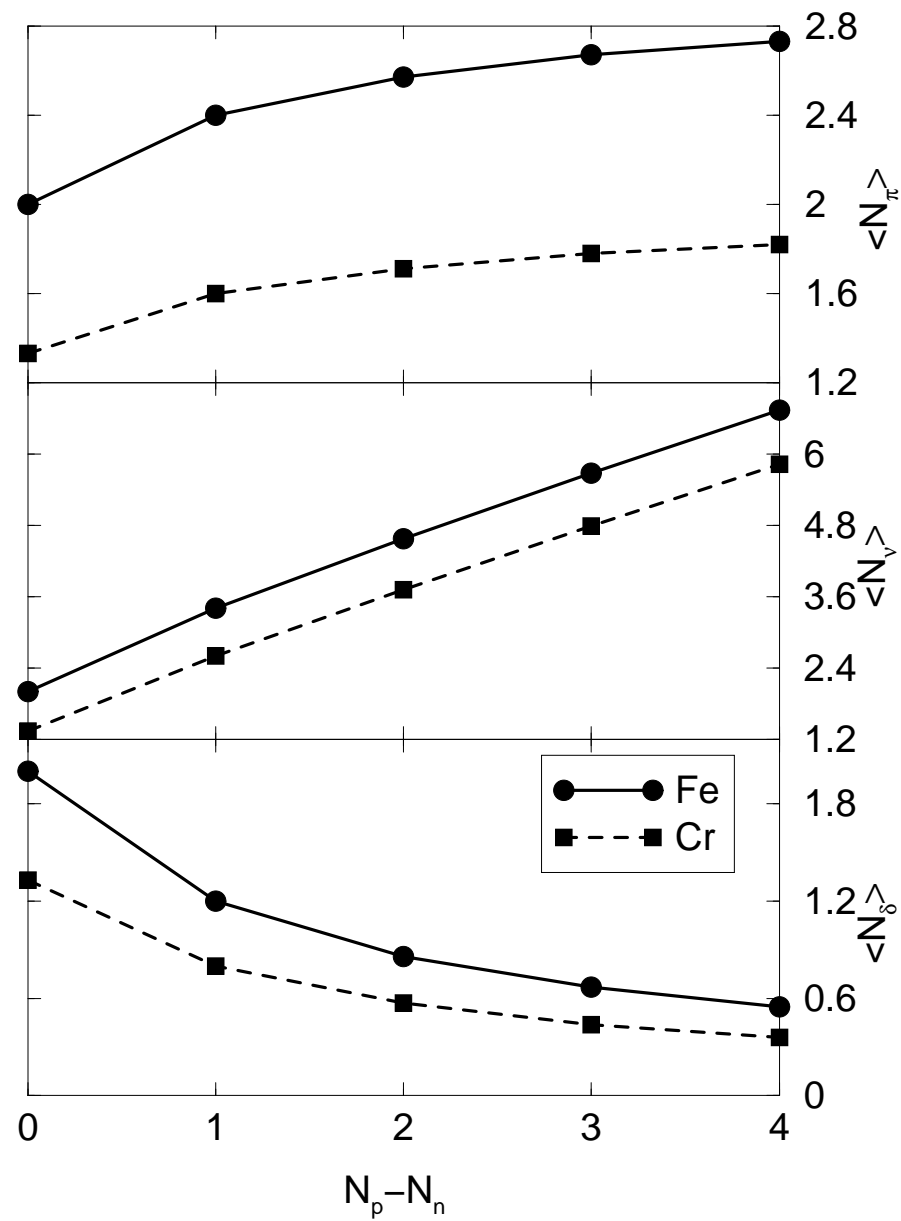
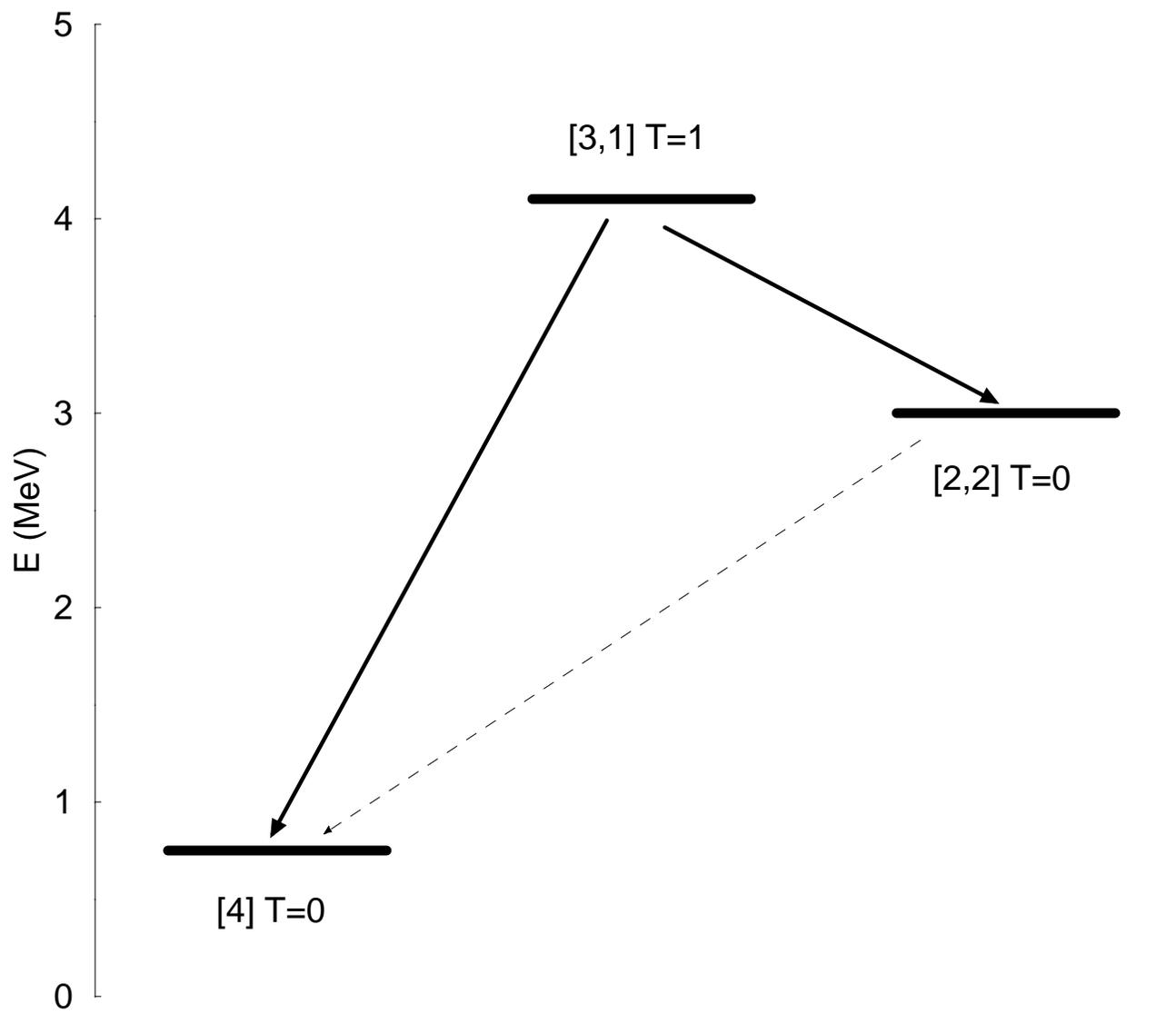
\end{document}